\documentclass[a4paper,twocolumn,showpacs,preprintnumbers,amsmath,amssymb]{revtex4}
\usepackage{amsmath,amssymb}
\usepackage[dvips]{graphicx}
\usepackage[dvips]{color}

\begin{document}

\title{Oscillatory Notch pathway activity in a delay model of neuronal differentiation}

\author{Hiroshi Momiji}
 \affiliation{Department of Biomedical Science,
   University of Sheffield,\\
   Western Bank, Sheffield S10 2TN, UK.}
 \email{h.momiji@sheffield.ac.uk}

\author{Nicholas A.M. Monk}
 \affiliation{School of Mathematical Sciences,
   University of Nottingham,\\
   University Park, Nottingham, NG7 2RD, UK.}
 \email{nick.monk@nottingham.ac.uk}

\date{\today}

\begin{abstract}
Lateral inhibition resulting from a double-negative feedback loop underlies the assignment of different fates to cells in many developmental processes. Previous studies have shown that the presence of time delays in models of lateral inhibition can result in significant oscillatory transients before patterned steady states are reached. We study the impact of local feedback loops in a model of lateral inhibition based on the Notch signalling pathway, elucidating the roles of intracellular and intercellular delays in controlling the overall system behaviour. The model exhibits both in-phase and out-of-phase oscillatory modes, and oscillation death. Interactions between oscillatory modes can generate complex behaviours such as intermittent oscillations. Our results provide a framework for exploring the recent observation of transient Notch pathway oscillations during fate assignment in vertebrate neurogenesis.
\end{abstract}




\maketitle

\section{Introduction}
As in many electronic circuits, classes of oscillators and switches
are fundamental elements in many gene regulatory networks
\cite{TysonJJ03,KruseK05}. In particular, a double-negative feedback
loop comprising two mutually repressive components  is known to be
capable of functioning as a toggle switch, allowing a system to
adopt one of two possible steady states (corresponding to cell
fates) \cite{CherryJL00, FerrellJE02}. In the context of
developmental biology, such bistable switch networks can operate
between cells, and are believed to drive cell differentiation in a
wide range of contexts. However, in naturally evolved (rather than
engineered) gene regulatory networks, double-negative feedback loops
rarely exist in a ``pure'' form, and interactions between loop
components and other network components often result in sets of
interconnected feedback loops. Furthermore, if loop interactions
involve the regulation of gene expression, then interactions are
delayed rather than instantaneous. The present study investigates
the dynamic behaviour of a double-negative feedback loop when the
nodes of the loop are involved in additional feedback loops, and
when the regulatory steps constituting the resulting network involve
significant time delays.

\begin{figure}[htbp] 
       \includegraphics[scale=0.4]{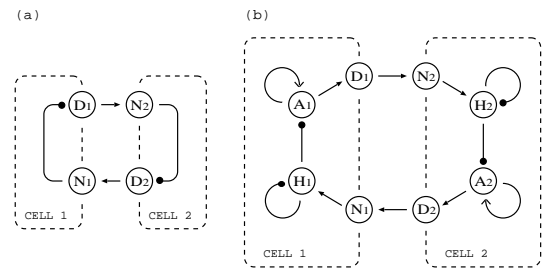} 
       \caption{\label{fig:delta_notch} The neurogenic network models in a two-cell
representation. The letters on the nodes correspond to the key
classes of protein involved in the network --- D: DSL ligand, N:
Notch receptor, H: Hes/Her proteins, A: proneural proteins (e.g.\
Achaete, Scute, Neurogenin). Edges represent two types of
directional interactions: activation ($\rightarrow$) and repression
($\multimap$). (a) The pure double-negative feedback loop involving
only DSL ligand and Notch receptor
\cite{CollierJR96,VeflingstadSR05}. (b) A more detailed model that
incorporates Hes/Her negative feedback and proneural positive
feedback \cite{MeirE02}.}
\end{figure}

A particularly well documented example of a biological
double-negative feedback loop is centred on transmembrane receptors
of the Notch family.  Notch signalling, resulting from direct
interaction with transmembrane ligands of the DSL (Delta, Serrate
and Lag-2) family on neighbouring cells mediates an
evolutionarily-conserved lateral inhibition mechanism that operates
to specify differential cell fates during many developmental
processes \cite{Artavanis-TsakonasS99, LewisJ98, LouviA06,
RichardsGS08}. Although gene nomenclature varies between different
organisms, a core circuitry---the neurogenic network---underlying
lateral inhibition can be identified, and is illustrated
schematically in Fig. \ref{fig:delta_notch} \cite{CollierJR96,
MeirE02}. In brief, signalling between neighbouring cells is
mediated by direct (juxtacrine) interactions between Notch receptors
and DSL ligands. A double-negative feedback loop is formed by
repression of DSL ligand activity by Notch signalling in the same
cell (cell autonomous repression)---Fig. \ref{fig:delta_notch}(a).
Mathematical models of such a spatially-distributed double-negative
feedback loop are capable of generating robust spatial patterns of
Notch signalling in populations of cells \cite{CollierJR96}.

In many developmental settings, the level of Notch signalling
regulates the fate adopted by a cell by acting as an input to a cell
autonomous bistable switch formed by one or more proneural genes
(such as {\em achaete} and {\em scute} in {\em Drosophila} and {\em
neurogenin} and {\em atonal} in vertebrates) \cite{BertrandN02}. The
basic principle underlying this switch is the ability of the protein
products of the proneural genes to positively regulate transcription
of proneural genes, resulting in a direct positive feedback loop. In
many systems, including the developing nervous system, Notch
signalling regulates the proneural switch via regulation of the
expression of proteins of the Hes/Her family. These proteins act as
transcriptional repressors, and can repress their own expression and
interfere with proneural self-activation \cite{KageyamaR07}.
Furthermore, proneural proteins can also positively regulate the
expression of DSL proteins, forming a complete circuit of
interactions as shown in Fig. \ref{fig:delta_notch}(b). Considered
as an intercellular signalling network---which we shall refer to as
the neurogenic network---this circuit comprises a
spatially-distributed double-negative feedback loop with additional
local positive and negative feedback loops.

A detailed mathematical model of the neurogenic network,
incorporating Hes/Her negative feedback and proneural positive
feedback, has been studied by Meir {\em et al.} \cite{MeirE02}, who
showed by computer simulation that the network is capable of
generating spatial patterns of Notch signalling in populations of
cells. The models of Collier {\em et al.} \cite{CollierJR96} and
Meir {\em et al.} \cite{MeirE02} incorporate the implicit assumption
that all interactions are non-delayed. However, in reality the basic
production mechanisms that regulate gene expression (gene
transcription and mRNA translation) are associated with time delays
\cite{MahaffyJM84, MahaffyJM88}. Incorporation of explicit time
delays in the pure double-negative feedback loop shown in Fig.
\ref{fig:delta_notch}(a) results in competition between dynamic
modes, with stable spatial patterning typically preceded by
significant oscillatory transients \cite{VeflingstadSR05}. In a
biological context, such transients would play an important role in
delaying the onset of cell differentiation in a developing tissue.
Delays can also generate oscillatory dynamics in models of Hes/Her
negative feedback loops \cite{JensenMH03, LewisJ03, MonkNAM03,
MomijiH08}, and such oscillations have been observed experimentally
\cite{HirataH02, MasamizuY06, KageyamaR07}.

As predicted by mathematical models \cite{LewisJ03,
VeflingstadSR05}, oscillatory expression of DSL ligands, Hes/Her
proteins and proneural proteins has recently been observed in neural
precursors in the developing mouse brain \cite{ShimojoH08}.
Furthermore, these oscillations have been predicted to play a
central biological role in delaying the onset of neural differentiation
\cite{ShimojoH08}. In principle, these oscillations could be driven
by the cell autonomous Hes/Her negative feedback loop, with Notch
signalling providing coupling between cells \cite{LewisJ03}, or by
the double-negative feedback loop centred on the DSL--Notch
interaction \cite{VeflingstadSR05}. In the following, we investigate
the interplay of local and intercellular feedback loops in models of
the neurogenic network, using a combination of linear stability
analysis and numerical simulations, emphasising the dynamical
effects of the multiple time delays in the network. We study principally
the case of two coupled cells, since this captures the
essential features of oscillator synchronisation and cell state
differentiation. We also show how our results extend to larger populations of coupled cells.

\section{The full Hes/Her--Proneural model and its dissection}
\begin{figure}[htbp] 
       \includegraphics[scale=0.4]{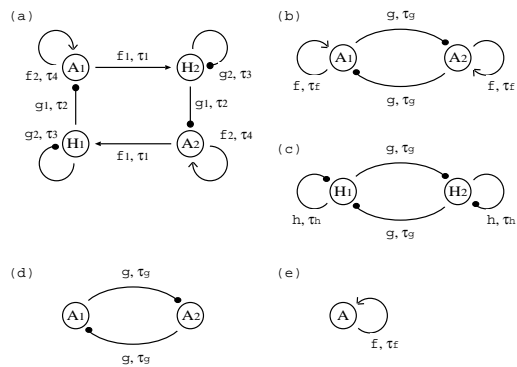} 
       \caption{\label{fig:motifs} The neurogenic network models examined in the present
study. $\tau$: time delays, $f$: a general increasing function; $g$,
$h$: general decreasing functions. With an appropriate choice of
time delay ($\tau_1$), the model in (a) provides a simplified
representation of the full model shown in Fig.
\ref{fig:delta_notch}(b). The models in (a)--(d) have the potential
to exhibit both oscillations and differentiation, while the model in
(e) exhibits only differentiation.}
\end{figure}

In Fig.~\ref{fig:delta_notch}(b), positive regulation of Hes/Her (H)
by proneural protein (A) in the adjacent cell, mediated by
DSL--Notch signaling, can be considered simply as a cascade of three
low-pass filters. This simplification allows the model in
Fig.~\ref{fig:delta_notch}(b) to be reduced to the model in
Fig.~\ref{fig:motifs}(a), where $\tau$ denotes time delay, and $f$
and $g$ represent generic increasing and decreasing functions,
respectively. In this model, referred to hereafter as the full
Hes/Her--Proneural model, Hes/Her proteins repress proneural
proteins in the same cell (1 or 2), while proneural proteins
activate Hes/Her proteins in the adjacent cell. This main loop is
supplemented by the two local loops: Hes/Her auto-repression and
proneural auto-activation. Each interaction is not instantaneous but
involves a time delay, typically of the order of minutes to tens of
minutes \cite{JensenMH03,LewisJ03,MonkNAM03,VeflingstadSR05}. The
delays in the cell-autonomous regulatory steps ($\tau_2$--$\tau_4$)
originate predominantly from processes associated with gene
transcription, whereby the DNA sequence of a gene is transcribed
into a corresponding mRNA molecule, while the delay in the
non-cell-autonomous interaction ($\tau_1$) represents in addition
processes involved in DSL--Notch signalling, such as protein
processing \cite{NicholsJT07}.

The models in Fig.~\ref{fig:motifs}(b)--(e) are obtained by
sequential reduction of the full neurogenic network. When Hes/Her
feedback is negligible, the full Hes/Her--Proneural model in (a)
becomes the model in (b); while when proneural feedback is
negligible, the model in (a) becomes the model in (c). When both
local loops are negligible, models (b) and (c) can be reduced to the
model in (d). Finally, because two sequential repressions function
as a net activation, the model in (d) can be further reduced to the
model in (e). Such simple network elements appear repeatedly in gene
regulatory networks (and possibly in other biological and
non-biological networks), and are examples of what are often called
``motifs'' \cite{AlonU07}. The models in Fig.
\ref{fig:motifs}(b)--(e) are called hereafter: (b) the
auto-activation two-Proneural model; (c) the auto-repression
two-Hes/Her model; (d) the non-autonomous Proneural model; (e) the
auto-activation single Proneural model.

\section{Mathematical representation of the full Hes/Her--Proneural model}
We represent Hes/Her and proneural proteins by a single variable in
each cell. To investigate the behaviour of a network quantitatively,
it is convenient to scale each variable such that it lies in the
range $[0, 1]$. The full Hes/Her--Proneural model (Fig.
\ref{fig:motifs}(a)) can then be represented by the following
differential equations with discrete delays:
\begin{eqnarray}
   T_H\ \dot{H}_1 & = & -H_1 + P_H(A_2(t-\tau_1),H_1(t-\tau_3)),
   \label{eq:4var_1}\\
   T_A\ \dot{A}_1 & = & -A_1 + P_A(H_1(t-\tau_2), A_1(t-\tau_4)),
   \label{eq:4var_2}\\
   T_H\ \dot{H}_2 & = & -H_2 + P_H(A_1(t-\tau_1), H_2(t-\tau_3)),
   \label{eq:4var_3}\\
   T_A\ \dot{A}_2 & = & -A_2 + P_A(H_2(t-\tau_2), A_2(t-\tau_4)),
   \label{eq:4var_4}
\end{eqnarray}
where $T_H$ and $T_A$ denote the degradation constants (the inverses of the linear degradation rates) of $H$ and
$A$, respectively \cite{MeirE02}. $P_H$ and $P_A$ are functions
representing the rate of production of $H$ and $A$ respectively. The
activating and repressive action of the proneural and Hes/Her
proteins is captured by the following constraints:
\begin{eqnarray}
   \frac{\partial P_H}{\partial A} &>& 0, \quad \frac{\partial P_H}{\partial H} < 0,
   \label{eq:pfun_1}\\
   \frac{\partial P_A}{\partial A} &>& 0, \quad \frac{\partial P_A}{\partial H} < 0.
   \label{eq:pfun_2}
\end{eqnarray}

For our analysis of this and the reduced models, we need assume no more about the production functions than the conditions (\ref{eq:pfun_1}) and (\ref{eq:pfun_2}). For numerical simulations, specific functional forms must be assumed, and we take these functions to be products of increasing ($f$) and decreasing ($g$ or $h$) Hill functions. Specifically, we assume that $P_I(A, H) = f_I(A)g_I(H)$ for $I =
A$ or $H$, where
\begin{eqnarray}
   f_I(x,K,\nu) & = & \frac{x^\nu}{K^\nu + x^\nu},
   \label{eq:Hill+}\\
   g_I(x,K,\nu) & = & \frac{K^\nu}{K^\nu + x^\nu},
   \label{eq:Hill-}
\end{eqnarray}
and $K$ and $\nu$ represent the scaled threshold and the Hill coefficient, respectively. However, the qualitative behaviour of the model solutions is preserved for other choices of production functions that satisfy conditions (\ref{eq:pfun_1}) and (\ref{eq:pfun_2}).

\begin{figure}[htbp] 
       \includegraphics[scale=0.33]{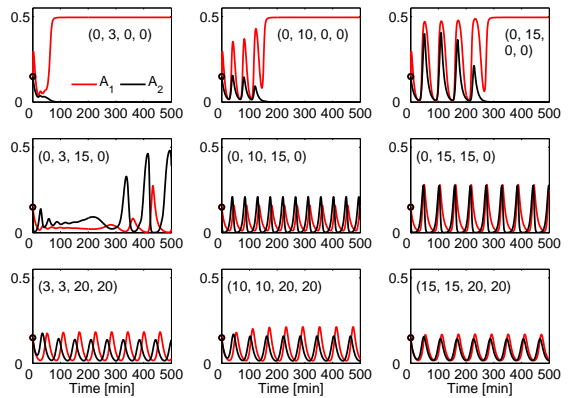} 
       \caption{\label{fig:sim_BABA} (Color online)
       Examples of typical dynamics found in numerical simulations of
       the full Hes/Her--Proneural model shown in Fig.\ \ref{fig:motifs}(a)
       and defined by Eqs.\ \ref{eq:4var_1}--\ref{eq:Hill-},
       displaying both differentiation and oscillations.
       Oscillations can be transient or sustained, and can be in-phase between two cells, or not.
       Values of the kinetic parameters are set the same on each row and listed in Table
       \ref{tbl:sim_BABA}. Delays are given in each panel as $(\tau_1, \tau_2, \tau_3, \tau_4)$}
\end{figure}

Numerical simulations of this model reveal a range of qualitatively
different types of behaviour, in which oscillations can be absent,
transient or sustained, and their phases can be locked or not (for
examples, see Fig. \ref{fig:sim_BABA}). To investigate the origin of
these behaviours, we reduce the full Hes/Her--Proneural model to a
variety of simpler ones (see Fig. \ref{fig:motifs}). The examination
of these simpler networks (motifs) helps to elucidate the origins of
the dynamics of the full Hes/Her--Proneural model.

\begin{table}
\caption{\label{tbl:sim_BABA} Parameter values.}
\begin{ruledtabular}
\begin{tabular}{ccccccccccc}
           &       & \multicolumn{2}{c}{$f_H$} & \multicolumn{2}{c}{$g_H$} &       & \multicolumn{2}{c}{$g_A$} & \multicolumn{2}{c}{$f_A$}\\
\cline{3-4} \cline{5-6} \cline{8-9} \cline{10-11}
           & $T_H$ & $K$       & $\nu$         & $K$       & $\nu$         & $T_A$ & $K$       & $\nu$         & $K$       & $\nu$\\
\hline
Standard    & 10 & 0.01 & 2 & 0.01 & 2 & 10 & 0.01 & 2 & 0.01 & 2\\
\hline
Fig. \ref{fig:sim_BABA}\\
Top         & 1  & 0.1  & 2 & 0.1  & 0 & 10 & 0.01 & 2 & 0.01 & 2\\
Middle      & 10 & 0.1  & 2 & 0.01 & 2 & 1  & 0.01 & 2 & 0    & 0\\
Bottom      & 10 & 0.01 & 2 & 0.01 & 2 & 10 & 0.01 & 2 & 0.01 & 2
\end{tabular}
\end{ruledtabular}
\end{table}

\section{Analysis and simulation of reduced models}
In this study, we are concerned primarily with the routes that cells
take to differentiation. For all model variants, a uniform steady
state exists ($H_1^*=H_2^*$, $A_1^*=A_2^*$). Biologically, this
corresponds to a non-differentiated state (i.e.\ the neurogenic network is in the same state in both cells). We therefore study the
stability of this steady state to small perturbations, and seek to
determine the resulting dynamical behaviour of the system in cases
where it is unstable.

For each model variant (Fig.~\ref{fig:motifs}(b)--(e)), we first
perform linear stability analysis of the uniform steady state, which
yields an eigenvalue equation. We then determine parameter values
that result in the existence of neutral (pure imaginary)
eigenvalues, which are the product of the imaginary unit number and the
neutral angular frequency, and are associated with changes in stability.
To confirm the nature of bifurcations associated with these
eigenvalues, the linear analysis is followed by numerical
simulations. The eigenvalue equation for each model variant is
derived directly from its own model equations, rather than by reduction of the full Hes/Her--Proneural model. The conditions under which
the full Hes/Her--Proneural model can be reduced to the
auto-activation two-Proneural model, and to the auto-repression
two-Hes/Her model are discussed in Sect.~\ref{sec:full}.

To allow systematic comparison between model variants, we use a standard set of parameter values---which are listed in Table
\ref{tbl:sim_BABA}---in both analysis and simulations. These
parameter values fall into a biologically reasonable range
\cite{MeirE02,LewisJ03} and result in representative model dynamics.
Deviations from this standard set are noted explicitly. Time is
measured in minutes, yielding values for the degradation constants
that are in line with measured values for proneural and neurogenic
factors \cite{HirataH02,VosperJMD07}. However, the behaviour of each model variant is also examined with other
parameter values.

In the following analyses, $\gamma$ denotes the {\em magnitude} of
the slope of the regulatory function ($f_{j=1,2}(x)$ or
$g_{j=1,2}(x)$) at the uniform steady-state solution ($H_{i=1,2}^*$
or $A_{i=1,2}^*$). In the cases of the Hill functions:
\begin{eqnarray}
   \gamma_{i=1,2}^{g_{j=1,2}} & = & -\frac{dg_{j=1,2}}{dx_i}|_{x_i^*}
       = \frac{\nu K^\nu {x_i^*}^{\nu-1}}{(K^\nu +
       {x_i^*}^\nu)^2} > 0,\\
   \gamma_{i=1,2}^{f_{j=1,2}} & = & +\frac{df_{j=1,2}}{dx_i}|_{x_i^*}
       = \frac{\nu K^\nu {x_i^*}^{\nu-1}}{(K^\nu +
       {x_i^*}^\nu)^2} > 0,
\end{eqnarray}
where, in general, $K$ and $\nu$ are different in $f_{j=1,2}(x)$ and
in $g_{j=1,2}(x)$, and therefore $\gamma_{i=1,2}^{g_{j=1,2}} \neq
\gamma_{i=1,2}^{f_{j=1,2}}$. In the following sections, it can be
seen that what determines the stability properties of the homogeneous steady state is not the  precise functional forms of $f$ and $g$, but the value of $\gamma$.

\subsection{Non-autonomous two-Proneural model}
The non-autonomous two-Proneural model (Fig. \ref{fig:motifs}(d)) is
represented by the following differential equations:
\begin{eqnarray}
   T \dot{A}_1 & = & -A_1 + g(A_2(t-\tau)),
   \label{eq:model_AA_no_1}\\
   T \dot{A}_2 & = & -A_2 + g(A_1(t-\tau)).
   \label{eq:model_AA_no_2}
\end{eqnarray}
To study the stability of the uniform steady state solution ($A_1^* = A_2^* = A^*$), we set $A_1(t) = A^* + a_1(t)$ and $A_2(t) = A^* + a_2(t)$ in Eqs. (\ref{eq:model_AA_no_1}) and
(\ref{eq:model_AA_no_2}). Following linearisation, the resulting coupled equations for $a_1$ and $a_2$ can be uncoupled by introducing variables $a_1+a_2$ and $a_1-a_2$ \cite{CollierJR96}.  The eigenvalue equations derived from the equations in these variables are:
\begin{equation}
   1 + T \lambda = \pm \gamma\ e^{-\lambda \tau},
   \label{eq:eigen_AA_no_auto}
\end{equation}
where the plus and minus branches are associated with $a_1 - a_2$ and $a_1 + a_2$ respectively.

\begin{figure}[htbp] 
       \includegraphics[scale=0.3]{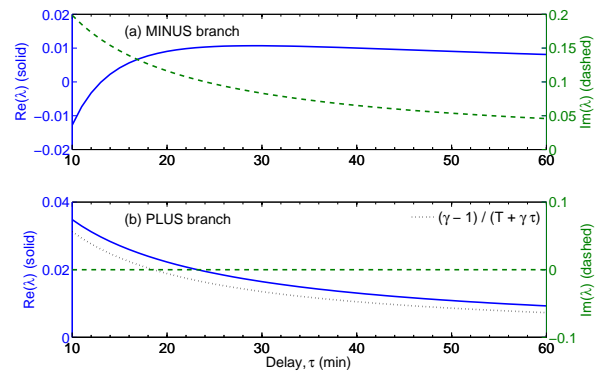} 
       \caption{\label{fig:lnr_AA_no} (Color online)
       The eigenvalues ($\lambda$) of the non-autonomous two-Proneural model (Fig.~\ref{fig:motifs}(d)) as a function of delay ($\tau$). The eigenvalue equation (Eq.~(\ref{eq:eigen_AA_no_auto})) has two branches. (a) On the minus branch $Re(\lambda)$ changes its sign from negative to positive as $\tau$ surpasses its critical value ($\tau_c = 13.0548$), leading to a Hopf bifurcation, (b) while on the plus branch it is always positive. The minus and plus branches therefore define respectively the oscillatory and differentiating properties of the system. The dotted line in (b) shows the approximate analytical value for the real part of the eigenvalue derived in Eq.~(\ref{eq:diff_speed_exponent}).}
\end{figure}

We consider first uniform perturbations such that $a_1 = a_2$. In this case, only the minus branch of Eq.~(\ref{eq:eigen_AA_no_auto}) is relevant.  Assuming a pure imaginary eigenvalue $\lambda = i \omega_c$, with $\omega_c$ real, the neutral angular frequency ($\omega_c$) is derived to be:
\begin{equation}
   \omega_c = \frac{1}{T} \sqrt{\gamma^2 - 1}.
\end{equation}
This eigenvalue occurs for parameter values such that $\gamma > 1$ and when the delay is equal to the neutral delay $\tau_c$:
\begin{equation}
   \tau_c = \frac{1}{\omega_c} \left(\pi + \tan^{-1} \left( -T \omega_c \right) \right).
\end{equation}
For the standard set of parameter values (Table I), the oscillatory period associated with the neutral eigenvalue (the Hopf period)---defined by $T_c = 2 \pi / \omega_c$---is found to be 38.6501 min, while $\tau_c = 13.0548$ min. For $\tau \neq \tau_c$, the minus branch of Eq.~(\ref{eq:eigen_AA_no_auto}) has a complex eigenvalue with a real part that has the same sign as $\tau - \tau_c$. This can be seen in the numerical solution of  Eq.~(\ref{eq:eigen_AA_no_auto}) in Fig.~\ref{fig:lnr_AA_no}.

\begin{figure}[htbp] 
       \includegraphics[scale=0.5]{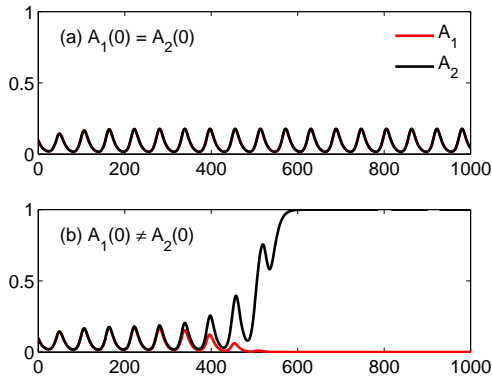} 
       \caption{\label{fig:sim_AA_no} (Color online)
       Numerical simulations of the non-autonomous two-Proneural model
(Fig. \ref{fig:motifs}(d)). (a) With identical initial values, $A_1$ and $A_2$ show sustained in-phase oscillations. (b) With slightly different initial values, $A_1$ and $A_2$ show transient in-phase oscillations followed by differentiation (cf.\ \cite{VeflingstadSR05}). Standard parameter values are used, as listed in Table 1, with $\tau = 20$ min.}
\end{figure}

The linear stability analysis suggests that the uniform steady state becomes unstable to small uniform perturbations via a Hopf bifurcation if $\gamma > 1$ and the delay increases above the critical value. Numerical simulations of Eqs. (\ref{eq:model_AA_no_1}) and (\ref{eq:model_AA_no_2}) for $\tau > \tau_c$ are shown in Fig.~\ref{fig:sim_AA_no}. For uniform initial conditions ($A_1(0) = A_2(0)$) the system exhibits sustained
oscillations (Fig.~\ref{fig:sim_AA_no}(a)). This confirms that the neutral solution on the minus branch of Eq.~(\ref{eq:eigen_AA_no_auto}) is a Hopf bifurcation point.

The plus branch of Eq.~(\ref{eq:eigen_AA_no_auto})---which is associated with $a_1 - a_2$---has the same neutral angular frequency ($\omega_c$) as the minus branch, and the neutral eigenvalue occurs for a neutral delay $\tau_c$:
\begin{equation}
   \tau_c = \frac{1}{\omega_c} \tan^{-1} \left(T \omega_c \right),
\end{equation}
which takes value 6.2702 min for the standard parameter set. More generally, the real part of a complex eigenvalue $\lambda_+$ satisfies
\begin{equation*}
   1 + T\ Re(\lambda_+) = \gamma\ e^{-Re(\lambda_+) \tau} \cos (Im(\lambda_+) \tau).
\end{equation*}
Since $Re(\lambda_+)$ is given by the intersection of $y = 1 + T Re(\lambda_+)$ and $y = \gamma \exp(-Re(\lambda_+) \tau)
\cos(Im(\lambda_+) \tau)$ on the $y$--$Re(\lambda_+)$ plane, a purely real $\lambda_+$ (for which $\cos (Im(\lambda_+) \tau) = 1$) has the largest real part, and is positive for all values of $\tau$ if $\gamma > 1$. This positive real eigenvalue is shown in Fig.~\ref{fig:lnr_AA_no}(b).  By assuming a purely real eigenvalue $(\lambda_{R+})$, and by applying the first-order Taylor expansion
to Eq. (\ref{eq:eigen_AA_no_auto}), a simple approximate (lower bound) expression for $\lambda_{R+}$ is derived to be
\begin{equation}
   \lambda_{R+} \simeq \frac{\gamma - 1}{T + \gamma \tau},
   \label{eq:diff_speed_exponent}
\end{equation}
which is confirmed to provide a good approximation to the growth
rate of patterns obtained in numerical simulations in Fig.~\ref{fig:lnr_AA_no}(b). Previously, such an approximation was
thought to be possible only in the limit of a large Hill coefficient
($\nu$) \cite{VeflingstadSR05}, while in Fig.~\ref{fig:lnr_AA_no},
$\nu = 2$.

The positive real eigenvalue on the plus branch corresponds to differentiation of the two cells (exponential growth of $a_1 - a_2$), while the complex eigenvalue associated with the minus branch corresponds to oscillations (in $a_1 + a_2$). Thus, if $A_1(0)$ and $A_2(0)$ are set slightly different, the behaviour of the system comprises a combination of these two fundamental modes. This can be seen in the numerical simulation of Eqs. (\ref{eq:model_AA_no_1}) and (\ref{eq:model_AA_no_2}) in Fig.~\ref{fig:sim_AA_no}(b), which shows a transient uniform oscillation followed by differentiation. Since the oscillations occur on the minus branch corresponding to $a_1 + a_2$, the oscillatory profiles of $A_1$ and $A_2$ are in-phase. Similar transient oscillations leading to differentiation have been observed previously in a delay model of the Delta-Notch signalling system (Fig.~\ref{fig:delta_notch}(a)) \cite{VeflingstadSR05}.

\begin{figure}[htbp] 
(a)        \includegraphics[scale=0.33]{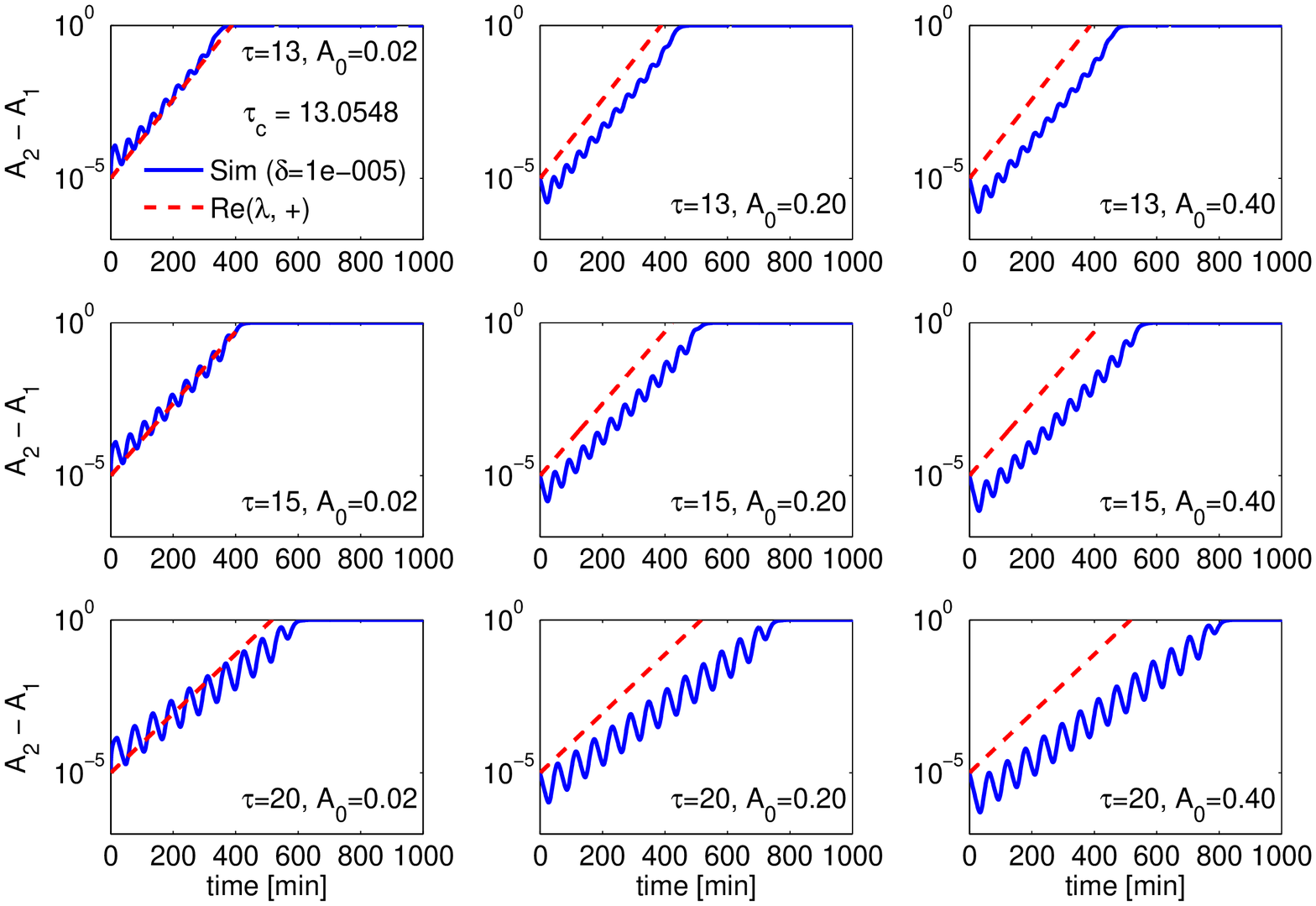}\\      
(b)       \includegraphics[scale=0.33]{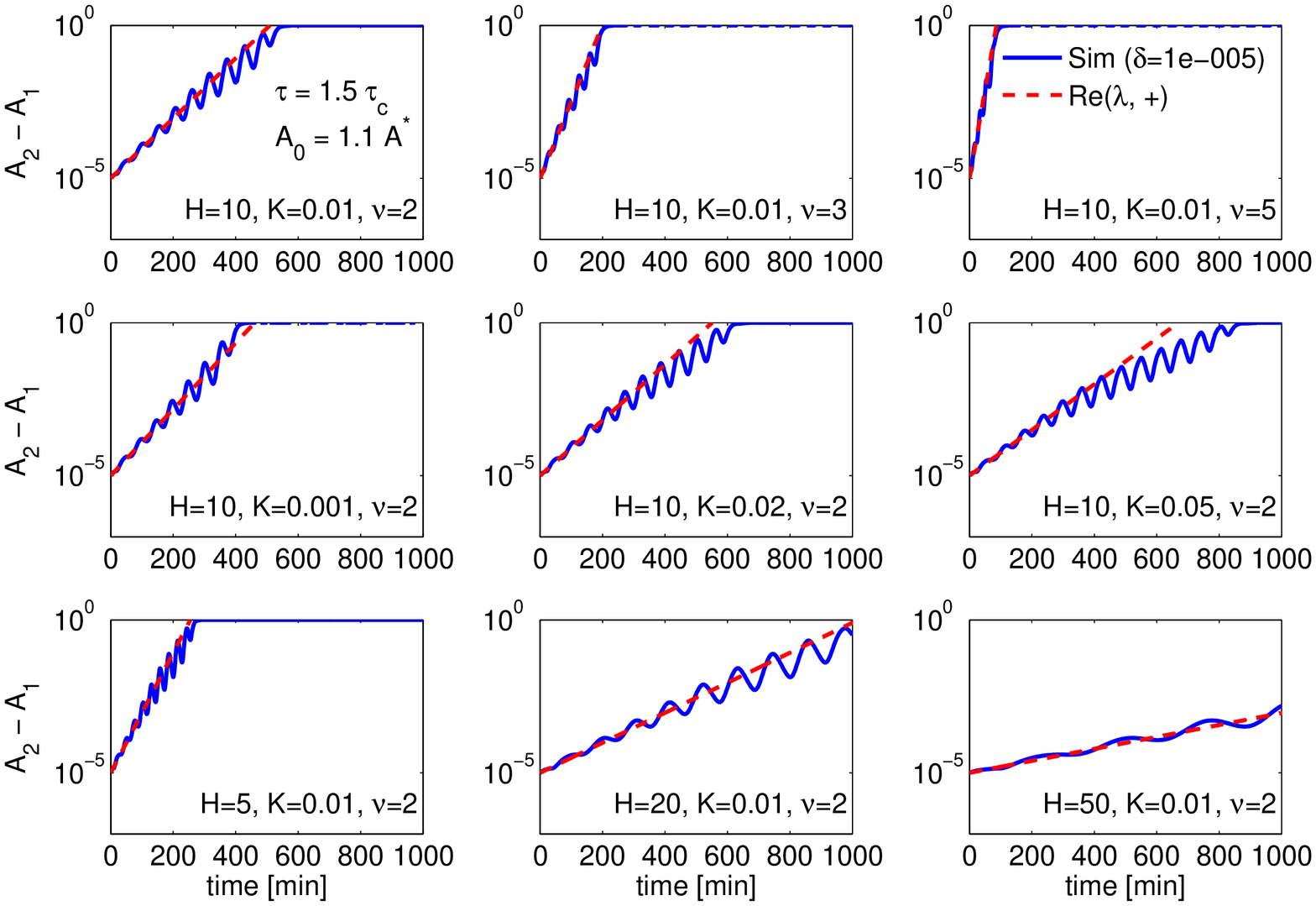}\\  
       \caption{\label{fig:diff_times} (Color online)
       Time taken for differentiation in the non-autonomous two-Proneural model (Fig.~\ref{fig:motifs}(d)). Time courses of the difference $A_2-A_1$---obtained from numerical simulations---plotted on a logarithmic scale, (a) for various delays ($\tau$) and initial mean values ($A_0 = 0.5\left[A_2(0) + A_1(0)\right]$) with the standard parameter set, and (b) for various parameter sets with $\tau = 1.5\ \tau_c (H,K,\nu)$, where $\tau_c$ is the critical delay. The time courses are compared to the growth rate predicted from linear analysis (dashed line),  based on the real part of the eigenvalue ($Re(\lambda)$).}
\end{figure}

In general, the outcome of linear stability analysis is not applicable to transient system behaviour. Therefore, the
separation of the two branches makes linear analysis highly
valuable in this study, making possible the prediction of whether or
not a transient oscillation leading to differentiation occurs.
More striking is the fact that not only the initial rate of differentiation,
but also the time taken to complete differentiation can be
estimated by linear analysis. Fig.~\ref{fig:diff_times} shows the
time-courses of the (logarithmic) difference $A_2 - A_1$ obtained
from numerical simulations for a range of values of the delay
($\tau$) and initial mean values ($A_0 = 0.5\left[A_2(0) + A_1(0)
\right]$). These are compared to the growth rate predicted by the
real part of the eigenvalue for $a_2 - a_1$ (i.e.\ the plus branch),
obtained from linear analysis for the corresponding $\tau$
(Fig.~\ref{fig:lnr_AA_no}), which provides a good estimate of the time taken to reach the fully differentiated state.

\subsection{Auto-activation single Proneural model}
Because two sequential repressions result in a net activation, the
non-autonomous two-Proneural model (Fig.~\ref{fig:motifs}(d)) may
seem to be surrogated by the auto-activation single Proneural model
(Fig.~\ref{fig:motifs}(e)), represented by the following
differential equation:
\begin{eqnarray}
   T \dot{A} & = & -A + f(A(t-\tau_f)).
   \label{eq:single_A}
\end{eqnarray}
The eigenvalue equation obtained by linearisation about a steady state of Eq.~(\ref{eq:single_A}), is
\begin{equation*}
   1 + T \lambda = \gamma\ e^{-\lambda \tau},
\end{equation*}
which is identical to the plus branch of
Eq.~(\ref{eq:eigen_AA_no_auto}) that represents the differentiating
nature of the non-autonomous two-Proneural model. It is therefore
clear that the oscillation in the two-Proneural model is due to the
network structure that two proneural proteins are mutually
repressing.

\subsection{Auto-activation two-Proneural model}
The auto-activation two-Proneural model (Fig.~\ref{fig:motifs}(b))
is represented by the following differential equations:
\begin{eqnarray}
   T \dot{A}_1 & = & -A_1 + g(A_2(t-\tau_g))\ f(A_1(t-\tau_f)),
   \label{eq:model_AA_1}\\
   T \dot{A}_2 & = & -A_2 + g(A_1(t-\tau_g))\ f(A_2(t-\tau_f)).
   \label{eq:model_AA_2}
\end{eqnarray}
The eigenvalue equation for the uniform steady-state solution
($A_1^*=A_2^*=A^*$) is derived to be:
\begin{equation}
   1 + T \lambda = g(A^*) \gamma^f e^{-\lambda \tau_f}
       \pm f(A^*) \gamma^g e^{-\lambda \tau_g}.
   \label{eq:eigen_AA}
\end{equation}
The purely imaginary solution $\lambda = i \omega$ satisfies:
\begin{equation}
   1+B^2-C^2+T^2 \omega^2 + 2 B (T \omega \sin \omega \tau_f - \cos \omega \tau_f) = 0,
   \label{eq:Hopf_AA}
\end{equation}
where $B = g(A^*) \gamma^f$ and $C = f(A^*) \gamma^g$.

\begin{figure}[htbp] 
       \includegraphics[scale=0.3]{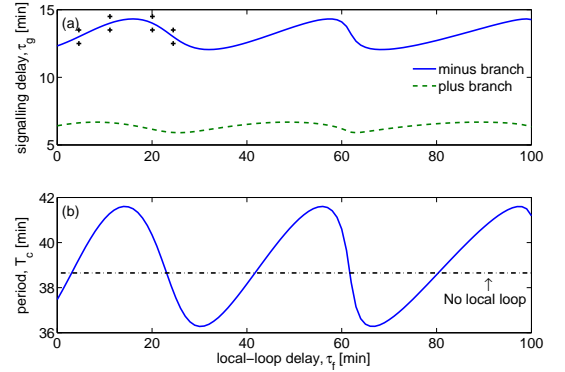} 
       \caption{\label{fig:lnr_AA} (Color online)
       Neutral intercellular signalling delay (a) and oscillatory period (b) associated with the pure imaginary eigenvalues of the auto-activation two-Proneural model (Fig.~\ref{fig:motifs}(b)), shown as a function of local-loop delay ($\tau_f$). The eigenvalue equation (Eq.~(\ref{eq:eigen_AA})) is solved numerically for its minus and plus branches. The Hopf bifurcation occurs on the minus branch. Also shown in (b) is the corresponding Hopf period for the non-autonomous two-Proneural model, which lacks the local (cell-autonomous) proneural feedback loops (Fig.~\ref{fig:motifs}(d)). Numerical
simulations performed for parameter values corresponding to the
cross signs in (a) confirm that the Hopf bifurcation occurs on the
minus branch of Eq.~(\ref{eq:eigen_AA}), and that the critical
intercellular signalling delay ($\tau_g$) is modulated by $\tau_f$.}
\end{figure}

Fig.~\ref{fig:lnr_AA} shows the neutral intercellular signalling
delay and corresponding oscillatory period, associated with the pure
imaginary eigenvalues, as a function of local-loop delay ($\tau_f$).
The eigenvalue equation (Eq.~(\ref{eq:eigen_AA})) is solved for its
minus and plus branches. As $\tau_f$ is varied, the values of the
neutral intercellular signalling delay and the neutral oscillatory
period fluctuate, with the oscillatory period fluctuating around its
value in the non-autonomous two-Proneural model
(Fig.~\ref{fig:motifs}(d)). A similar modulation in
period was recently observed in a delayed coupling model of
vertebrate segmentation \cite{MorelliLG09}. Numerical
simulations performed for the parameter values represented by the
cross signs in Fig.~\ref{fig:lnr_AA}(a) confirm that the Hopf
bifurcation occurs on the minus branch of Eq.~(\ref{eq:eigen_AA}),
and that the critical intercellular signalling delay is modulated by
the local-loop delay (data not shown).

\begin{figure}[htbp] 
       \includegraphics[scale=0.33]{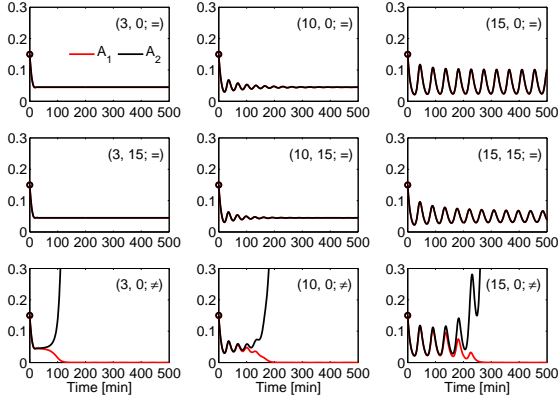} 
       \caption{\label{fig:sim_AA} (Color online)
       Numerical simulations of the auto-activation two-Proneural model (Fig.~\ref{fig:motifs}(b)).
       In comparison to Fig.~\ref{fig:sim_AA_no}, the behaviour is found to be qualitatively the same as that of the non-autonomous two-Proneural model (Fig.~\ref{fig:motifs}(d)).
       However, the modulatory effect of the cell-autonomous proneural loop, shown in Fig.~\ref{fig:lnr_AA}, is seen in the comparison between top and the middle right panels.
       As the local-loop delay ($\tau_f$) increases from 0 to 15, the amplitude of oscillations decreases, influenced by the increase of the critical signalling delay (Fig.~\ref{fig:lnr_AA}(a), solid curve).
       Values of the delays, and whether or not the initial values of $A_1$ and $A_2$ are equal, are specified in each panel as $(\tau_g, \tau_f; = \text{ or } \neq)$.}
\end{figure}

Fig.~\ref{fig:sim_AA} shows the results of numerical simulations
of the model equations (\ref{eq:model_AA_1}) and
(\ref{eq:model_AA_2}) for a range of delays ($\tau_g$ and $\tau_f$)
and initial values ($A_1(0)$ and $A_2(0)$). The behaviour of this
model variant is found to be qualitatively the same as that of the
non-autonomous two-Proneural model (Fig.~\ref{fig:motifs}(d)): the
Hopf bifurcation point exists on the minus branch of the eigenvalue
equation (Eq.~(\ref{eq:eigen_AA})) and, since the plus branch again has a positive real eigenvalue, differentiation occurs when
$A_1(0) \neq A_2(0)$. However, the modulatory effect of the cell-autonomous proneural
auto-activation loop, shown in Fig.~\ref{fig:lnr_AA}, is seen in the
comparison between top and the middle right panels. As the
local-loop delay ($\tau_f$) increases from 0 to 15, the amplitude of
oscillations decreases, influenced by the increase of the critical signalling
delay (Fig.~\ref{fig:lnr_AA}(a), solid curve). It can be seen from
Fig.~\ref{fig:lnr_AA} that the cell-autonomous proneural auto-activation loops
give rise to `tunability' of the oscillations.

\subsection{Auto-repression two-Hes/Her model}
The auto-repression two-Hes/Her model (Fig.~\ref{fig:motifs}(c)) is
represented by the following differential equations:
\begin{eqnarray}
   T \dot{H}_1 & = & -H_1 + g(H_2(t-\tau_g))\ h(H_1(t-\tau_h)),
   \label{eq:model_BB_1}\\
   T \dot{H}_2 & = & -H_2 + g(H_1(t-\tau_g))\ h(H_2(t-\tau_h)).
   \label{eq:model_BB_2}
\end{eqnarray}

The eigenvalue equation for the uniform steady-state solution
($H_1^*=H_2^*=H^*$) has the same form as that for the
auto-activation two-Proneural model (Eq.~(\ref{eq:eigen_AA})):
\begin{equation}
   1 + T \lambda = -g(H^*) \gamma^h e^{-\lambda \tau_h}
       \pm h(H^*) \gamma^g e^{-\lambda \tau_g},
   \label{eq:eigen_BB}
\end{equation}
where, however, the definition of $B$ is modified to be $B = -g(H^*)
\gamma^h$ because $h$ represents a decreasing Hill function.

\begin{figure}[htbp] 
       \includegraphics[scale=0.3]{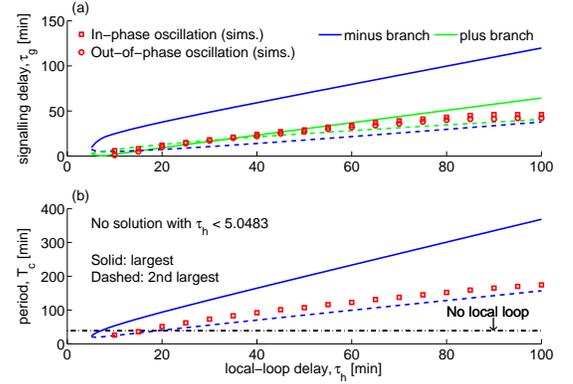} 
       \caption{\label{fig:lnr_BB} (Color online) Neutral intercellular signalling delay (a) and oscillatory period (b) associated with the pure imaginary eigenvalues of the auto-repression two-Hes/Her model (Fig.~\ref{fig:motifs}(c)), shown as a function of the local-loop delay. The eigenvalue equation (Eq.~(\ref{eq:eigen_BB})) is solved for its minus and plus branches, for the first and the second largest periods (solid and dashed lines, respectively). This model shows the transition between out-of-phase and in-phase
       oscillations. At each local-loop delay, a square represents the in-phase
       oscillation observed in simulation with the lowest signalling delay, while a circle represents the
       out-of-phase oscillation observed in simulation with the highest signalling delay.
       The transition happens between them (upper panel), which are overlapping with a square being always above a circle.}
\end{figure}

Fig.~\ref{fig:lnr_BB} shows the neutral intercellular signalling delay and oscillatory period associated with the
pure imaginary eigenvalues, as a function of the delay in the local
loop. The eigenvalue equation (Eq.~(\ref{eq:eigen_BB})) is solved
for its minus and plus branches, for the first and the second
largest periods. Unlike any eigenvalue equation of the three simpler
models analysed so far (Fig.~\ref{fig:motifs}(b), (d) and (e)),
Eq.~(\ref{eq:eigen_BB}) is found not to have a neutral solution when
$\tau_h < 5.0483$ min.

\begin{figure}[htbp] 
       \includegraphics[scale=0.33]{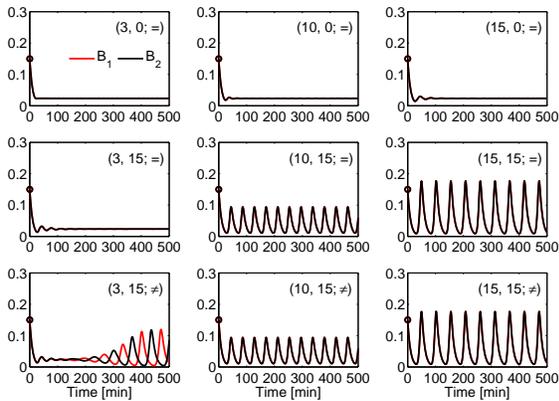} 
       \caption{\label{fig:sim_BB} (Color online) Numerical simulations of the auto-repression two-Hes/Her model (Fig.~\ref{fig:motifs}(c)). In comparison to Fig.~\ref{fig:sim_AA_no},
       the behaviour is found to be qualitatively different to that of the non-autonomous two-Proneural model (Fig.~\ref{fig:motifs}(d)).
       Specifically, oscillations are sustained even with non-identical initial values, and oscillations can be out-of-phase as well as in-phase.
       Values of the delays, and whether or not the initial values of $H_1$ and $H_2$ are equal, are specified in each panel as $(\tau_g, \tau_h; = \text{ or } \neq)$.}
\end{figure}

\begin{figure}[htbp] 
       \includegraphics[scale=0.33]{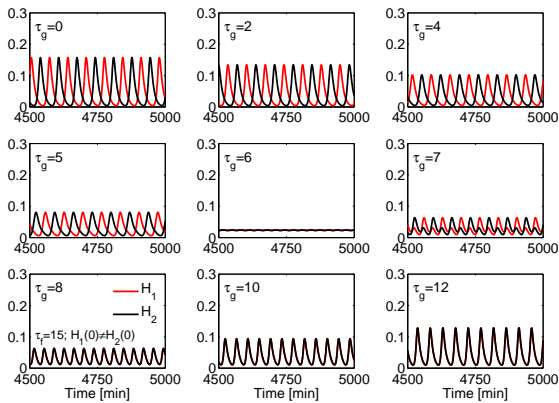} 
       \caption{\label{fig:mode_trans_1st} (Color online)
       Detail of the transition from out-of-phase to in-phase oscillations as the value of $\tau_g$ increases in the auto-repression two-Hes/Her model (Fig.~\ref{fig:motifs}(c); Fig.~\ref{fig:sim_BB}). The transition is associated with amplitude death around $\tau_g = 6$.}
\end{figure}

Numerical simulations of the model equations (\ref{eq:model_BB_1})
and (\ref{eq:model_BB_2}) reveal two prominent features of the
dynamics of this system that are qualitatively different to those of
the two-Proneural models (Figs.~\ref{fig:motifs}(b),(d)).
Fig.~\ref{fig:sim_BB} shows simulation results for a range of values
of the delays ($\tau_g$ and $\tau_h$) and initial values ($H_1(0)$
and $H_2(0)$). Two new features are apparent: first, oscillations
can be sustained (rather than transient) even when $H_1(0) \neq
H_2(0)$; second, the oscillations of $H_1$ and $H_2$ can be
out-of-phase for certain values of the delays, as shown in the left
bottom panel. Fig.~\ref{fig:mode_trans_1st} details the transition
from out-of-phase oscillations to in-phase oscillations as the value
of the critical intercellular signalling delay ($\tau_g$) increases.
Strikingly, the transition is found to be associated with a
point-wise amplitude death \cite{ReddyDVR98}. The transition found
in numerical simulations with various $\tau_h$ is compared to the
neutral properties estimated by the linear analysis in
Fig.~\ref{fig:lnr_BB}. In a cell-autonomous Hes/Her oscillator, the
oscillatory period increases monotonically with delay
\cite{MonkNAM03}, while for the coupled cells to cycle out-of-phase,
the signalling delay must be about half a period. Therefore the
monotonic increase in  the critical $\tau_g$ and in the critical period ($T_c$) suggests
that the overall network behaviour of the auto-repression
two-Hes/Her model is controlled by the two local autonomous Hes/Her
oscillators.

\begin{figure}[htbp] 
  \begin{center}
       \includegraphics[scale=0.4]{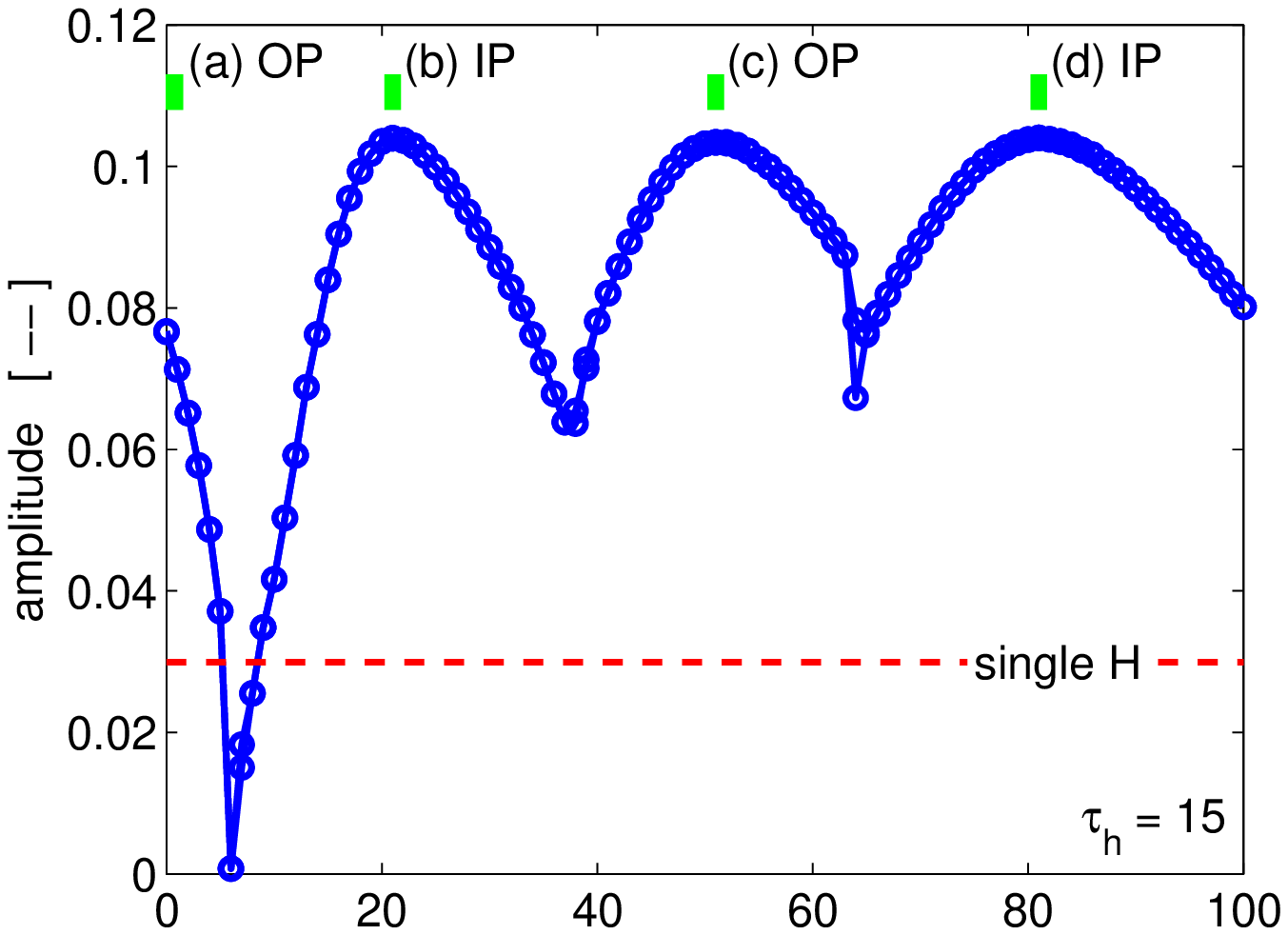}
      \\[-3mm]
       \includegraphics[scale=0.4]{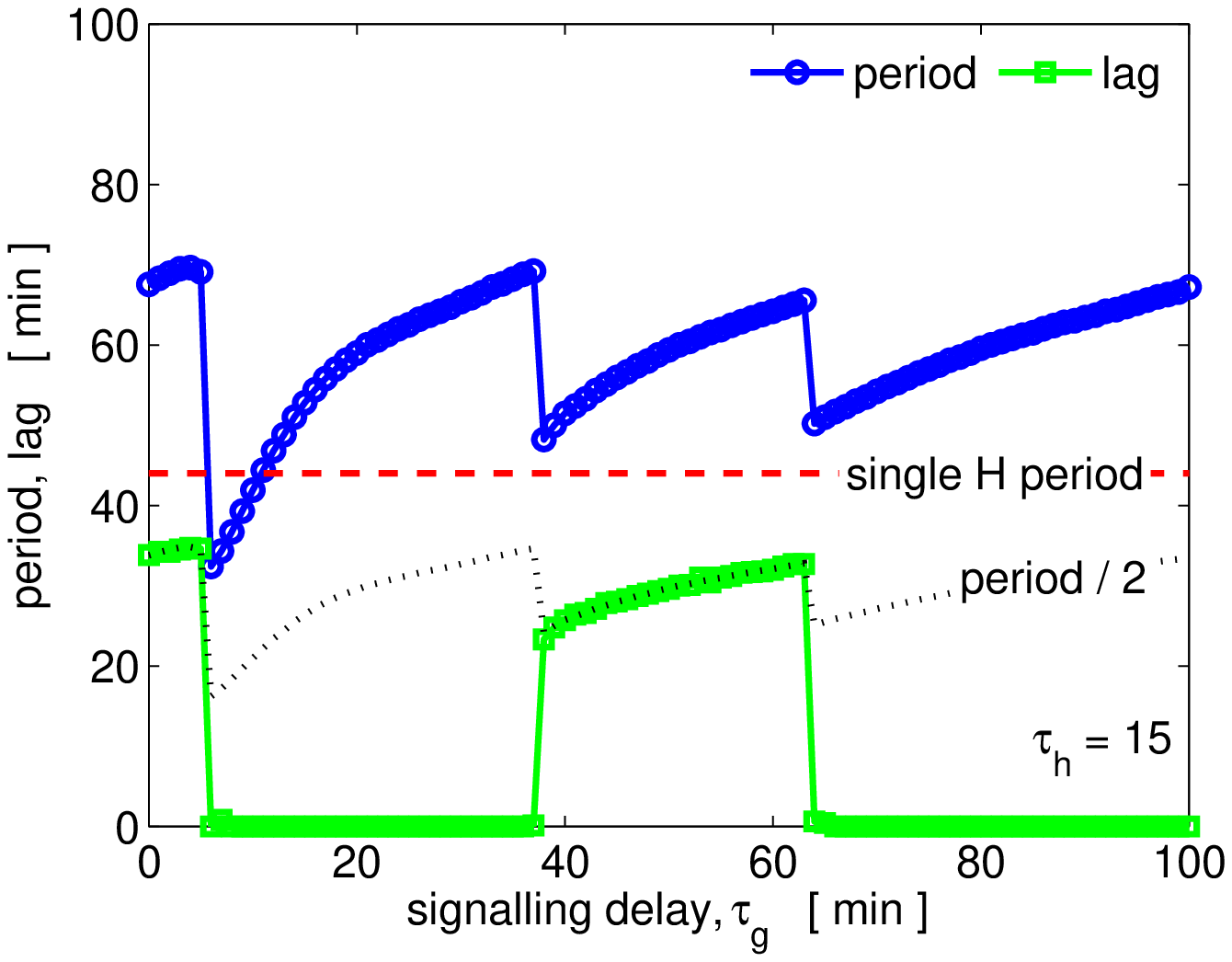}\\
       \includegraphics[scale=0.4]{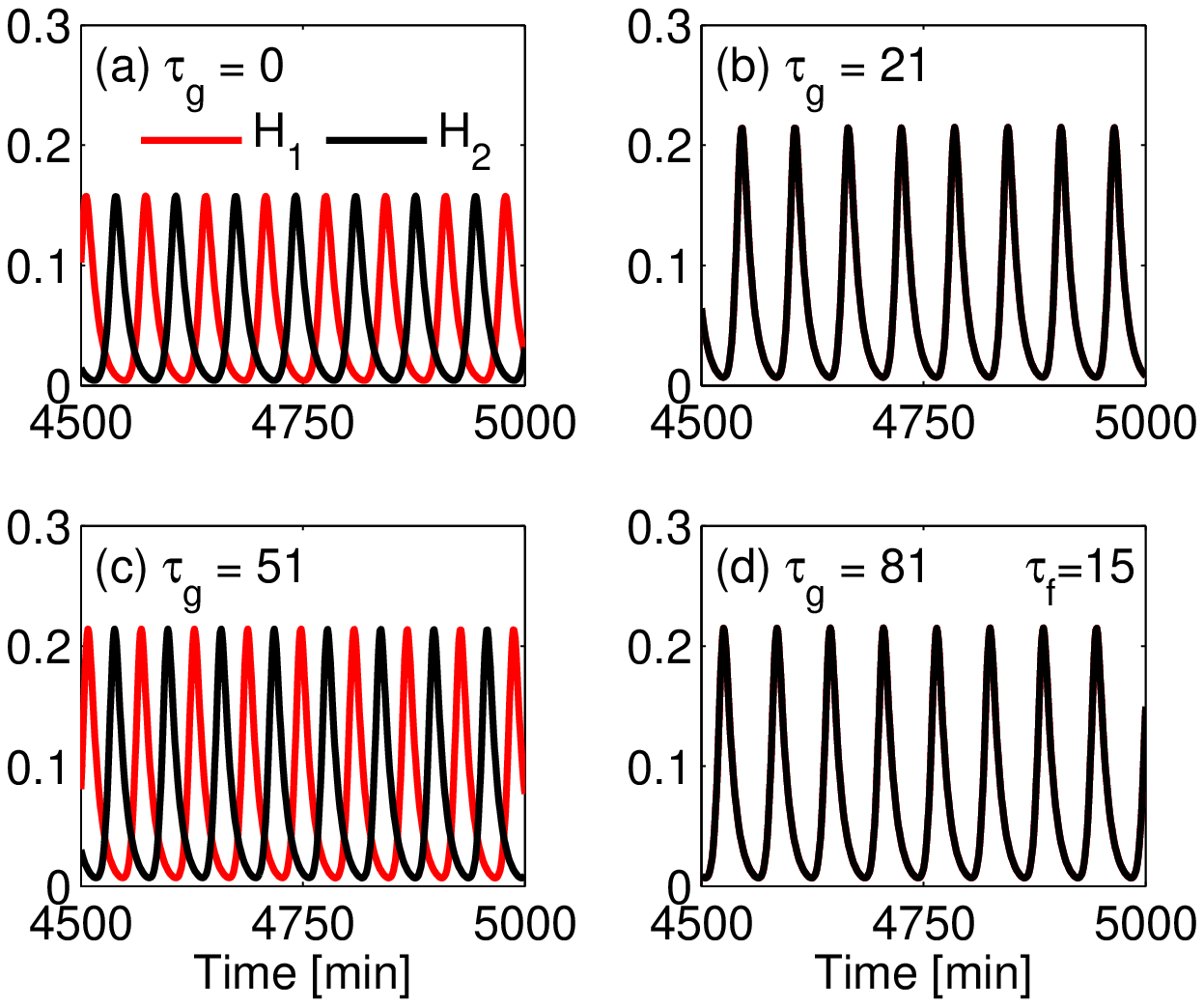}           
       \caption{\label{fig:Amp_Per_tg} (Color online)
       Dependence of the amplitude (top panel) and period (middle panel) of oscillatory solutions on the intercellular signalling delay ($\tau_g$) in the auto-repression two-Hes/Her model. The amplitude of an oscillatory solution is defined as the half difference between peak and trough values. The observed modulations of amplitude and period reflect the transition between out-of-phase (OP) and in-phase (IP) oscillation modes; the comparison between lag (time difference between oscillation peaks in neighbouring cells) and the half-period of oscillations shown in the middle panel illustrates the sequential switching between out-of-phase and in-phase oscillation modes. The bottom four panels show time courses obtained by numerical simulation of Eqs.~(\ref{eq:model_BB_1}) and (\ref{eq:model_BB_2}) at the four values of $\tau_g$ marked in the top panel. The first transition corresponds to Fig.~\ref{fig:mode_trans_1st}. The dashed lines in the top and the middle panels represent
respectively the amplitude and the period in the single-cell Hes/Her model with local delay $\tau_h = 15$.}
   \end{center}
\end{figure}

In contrast, when the intercellular signalling delay ($\tau_g$) is
varied while the local delay ($\tau_h$) is kept constant,
transitions between in-phase and out-of-phase modes occur
sequentially, and are associated with modulations in amplitude and
period (Fig.~\ref{fig:Amp_Per_tg}). The mode transitions repeat as
$\tau_g$ is increased, but not in an identical manner. Unlike the
first transition shown in Fig.~\ref{fig:mode_trans_1st}, a clear
amplitude death is not observed at higher-order transitions. Further
examination reveals that the sequential transitions are not induced
by the network structure of two mutually repressive autonomous
oscillators, but by the structure of two mutually influencing
autonomous oscillators, forming an overall positive feedback loop.
Indeed, when intercellular interactions are modelled by an
increasing Hill function that represents activation, the transition
still occurs, but in a reverse order, starting from the in-phase
mode at $\tau_g = 0$ (data not shown).

\begin{figure}[htbp] 
   \begin{center}
       \includegraphics[scale=0.7]{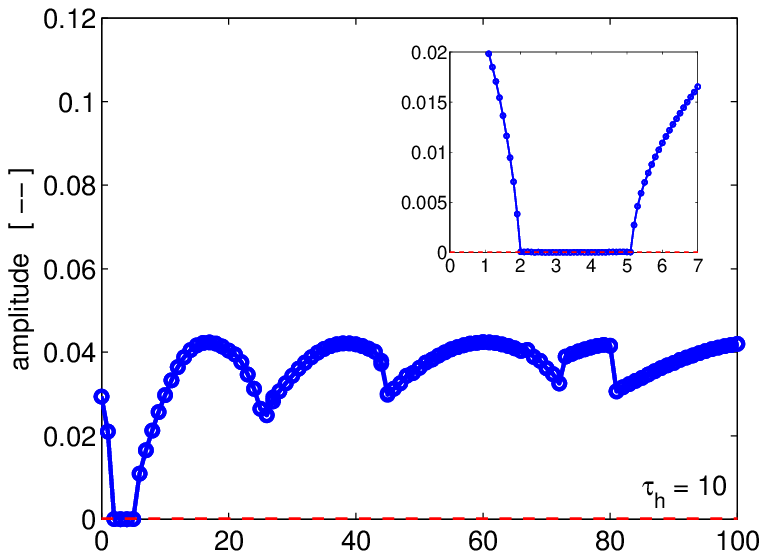}\\
       \includegraphics[scale=0.7]{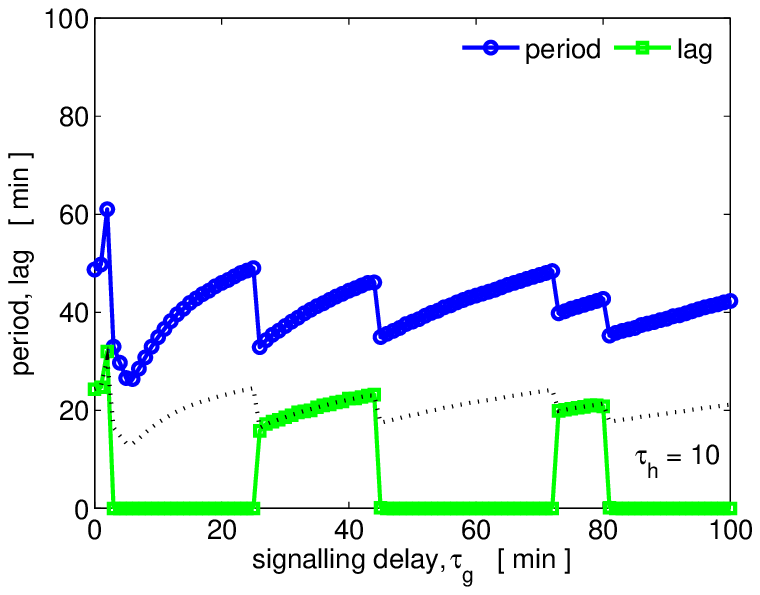}
       \caption{\label{fig:Amp_Per_tg_sub} (Color online)
       Coupled sub-critical oscillators illustrating a finite-$\tau_g$-range oscillation death. The inset in the top panel shows the oscillation death in detail.}
   \end{center}
\end{figure}

The intercellular coupling is found to have additional dynamical
consequences. While an isolated cell containing the Hes/Her feedback
loop cannot exhibit sustained oscillations when the delay ($\tau_h$)
is below its critical value, such a local-loop critical delay
($\tau_{h\ cr}$) can be lowered by the intercellular coupling, as
shown in Fig.~\ref{fig:Amp_Per_tg_sub}, where $\tau_h = 10 <
\tau_{h\ cr} = 13.0548$. Furthermore, oscillation death is found to
happen in a definite range of the intercellular signalling delay
($\tau_g$).

\begin{figure}[htbp] 
   \begin{center}
       \includegraphics[scale=0.5]{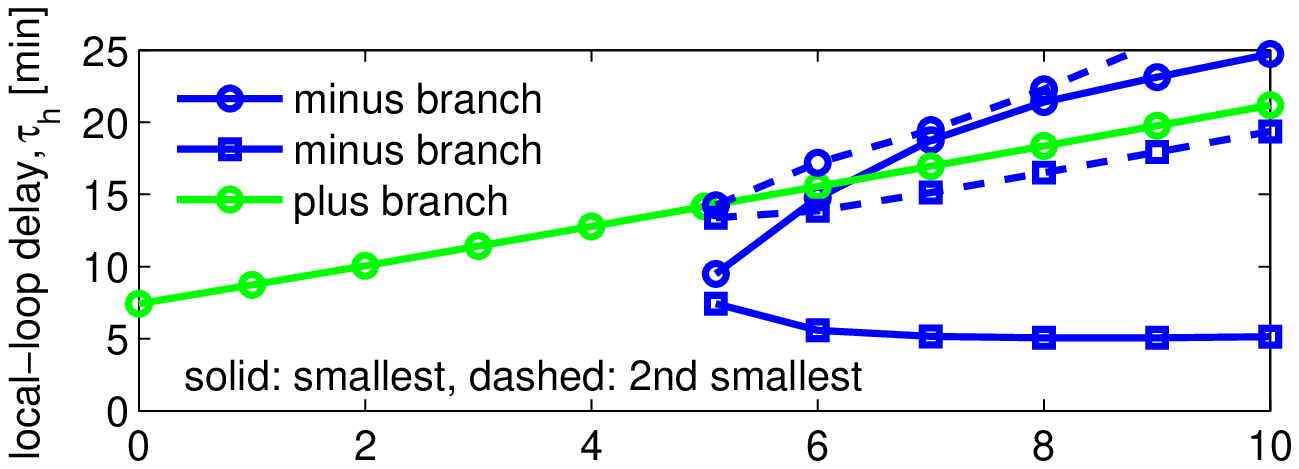}\\
       \includegraphics[scale=0.5]{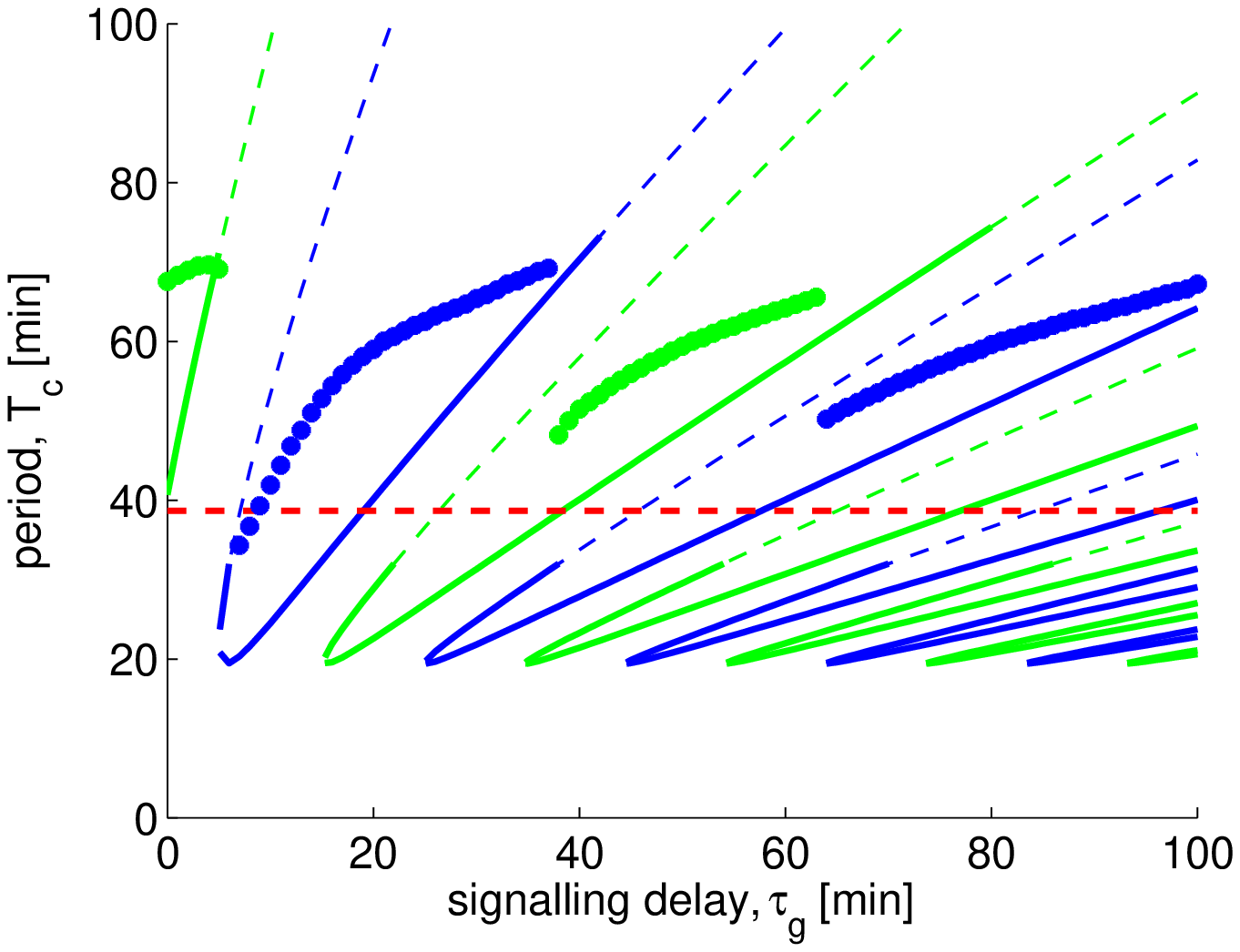}
       \caption{\label{fig:Anal_Oscil_Death} (Color online)
       The origin of oscillation death, and the transition between the out-of-phase and the in-phase oscillatory modes in the auto-repression two-Hes/Her model (Fig.~\ref{fig:motifs}(c)). Neutral local-loop delay (upper panel) and oscillatory period (lower panel) are shown as functions of the intercellular signalling delay $\tau_g$. The eigenvalue equation (Eq.~(\ref{eq:eigen_BB})) is solved numerically for its minus and plus branches. Neither the out-of-phase branch (green) nor the in-phase branch (blue) have a solution for $\tau_h = 10$ for $2 \leq\tau_g < 5.1$, corresponding to the oscillation death observed in Fig.~\ref{fig:Amp_Per_tg_sub}. Simulated periods for $\tau_h = 15$ (shown also in Fig.~\ref{fig:Amp_Per_tg}, middle panel) are shown by blue and green dots in the lower panel, where the neutral period obtained by linear stability analysis is shown by solid and dashed curves (correspond to the neutral delay being smaller or larger than $\tau_h = 15$, respectively). Note the different horizontal scales in the two panels.}
   \end{center}
\end{figure}

The three features observed in numerical simulations---the
oscillation in an originally subcritical delay range ($\tau_h <
\tau_{h\ cr}$), the transition between out-of-phase and in-phase
oscillatory modes, and the finite-$\tau_g$-range oscillation
death---can be understood analytically as follows. Linearisation of
Eqs.~(\ref{eq:model_BB_1}) and (\ref{eq:model_BB_2}) around the
uniform steady state yields:
\begin{equation}
\left( \begin{array}{cc}
1 + T \lambda + \alpha & \beta \\
\beta & 1 + T \lambda + \alpha
\end{array} \right)
\left( \begin{array}{c}
\bar{H}_1 \\
\bar{H}_2 \end{array} \right) = \mathbb{M} \left( \begin{array}{c}
\bar{H}_1 \\
\bar{H}_2 \end{array} \right) = \left( \begin{array}{c}
0 \\
0 \end{array} \right),
   \label{eq:eigen_BB_mat}
\end{equation}
where $\alpha = g(H^*) \gamma^h e^{-\lambda \tau_h}$, $\beta =
h(H^*) \gamma^g e^{-\lambda \tau_g}$. The associated eigenvalue
equation, given by $\det \mathbb{M} = 0$, is:
\begin{equation}
   1 + T \lambda = -\alpha \pm \beta,
   \label{eq:eigen_BB_alpha_beta}
\end{equation}
which is the same as Eq.~(\ref{eq:eigen_BB}). By substituting
Eq.~(\ref{eq:eigen_BB_alpha_beta}) into Eq.~(\ref{eq:eigen_BB_mat}),
it can be seen that the plus branch is associated with the
out-of-phase oscillation mode: $(\bar{H}_1, \bar{H}_2) = (1, -1)$,
whereas the minus branch is associated with the in-phase oscillation
mode: $(\bar{H}_1, \bar{H}_2) = (1, 1)$. Numerical evaluation of the
eigenvalue equation yields the relations between the neutral delay
in the local loop ($\tau_{h\ cr}$) and the intercellular signalling
delay ($\tau_g$) shown in Fig.~\ref{fig:Anal_Oscil_Death}, for both
the out-of-phase (green) and the in-phase (blue) modes. For example,
for $\tau_h = 10$ (Fig.~\ref{fig:Amp_Per_tg_sub}), when $0 < \tau_g
< 2$ only the out-of-phase mode is allowed, whereas when $\tau_g
\geq 5.1$ only the in-phase mode is allowed. In between ($2 < \tau_g
< 5.1$), no oscillation is allowed, explaining the origin of the
oscillation death shown in Fig.~\ref{fig:Amp_Per_tg_sub}.
However, the comparison of periods to simulation
results in the lower panel shows that as $\tau_g$ increases, more and
more branches appear, and the model behaviour becomes more
non-linear. The actual transitions observed in simulations do not coincide with the
boundaries between the regions that linear stability analysis predicts to
be competent (solid) and incompetent (dashed) for sustained
oscillations.

It is striking that only by changing the local loop attached to the
non-autonomous two-X model from auto-activation
(Fig.~\ref{fig:motifs}(b) where X is Proneural) to auto-repression
(Fig.~\ref{fig:motifs}(c) where X is Hes/Her), the network behaviour
changes completely. This simple example, motivated by the structure
of the neural differentiation network, highlights the importance of
examining in detail the structure of any gene regulatory network,
even if it is centred on a small and seemingly simple motif.

\begin{figure}[htbp] 
       \includegraphics[scale=0.33]{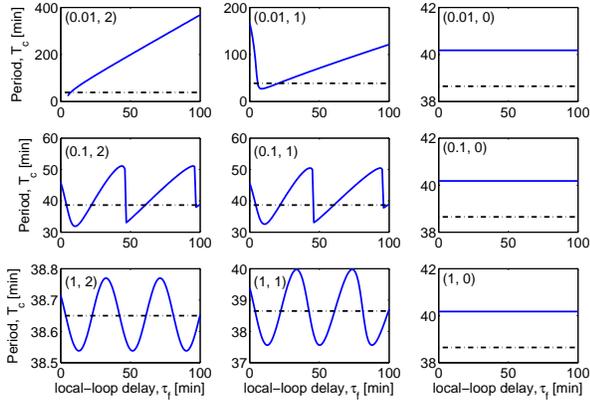} 
       \caption{\label{fig:lnr_BB_9pics} (Color online)
       The largest neutral period ($T_c$) associated with the eigenvalues of the auto-repression two-Hes/Her model (Fig.~\ref{fig:motifs}(c)) for different values of the threshold ($K_h$) and the
Hill coefficient ($\nu_h$) in the local feedback loop. As $K_h$
increases, the behaviour of this model becomes similar to that of
the auto-activation two-Proneural model (Fig.~\ref{fig:motifs}(b)),
while as $\nu_h$ decreases, the behavour becomes similar to that of
the non-autonomous two-Proneural model (Fig.~\ref{fig:motifs}(d)).
Note the different vertical scales in each plot. The dashed-dot line
in each panel represents the critical period in the non-autonomous
two-cell model. Varied parameters are given in each panel as ($K_h$,
$\nu_h$).}
\end{figure}

Richness in the behaviour of this auto-repression two-Hes/Her model
can be seen in Fig.~\ref{fig:lnr_BB_9pics}, where the largest
neutral period ($T_c$) associated with the eigenvalues is plotted
for different values of thee threshold ($K_h$) and Hill coefficient
($\nu_h$) in the local feedback loop. As $K_h$ increases, the
behaviour of this model becomes similar to that of the
auto-activation two-Proneural model (Fig.~\ref{fig:motifs}(b)),
while as $\nu_h$ decreases, the behavour becomes similar to that of
the non-autonomous two-Proneural model (Fig.~\ref{fig:motifs}(d)).

\section{Extension to $N$-cell ring model} \label{sec:ring}
The two-cell models discussed so far can be naturally extended to one-dimensional arrays of cells. To avoid potential boundary effects, we consider the specific case of a ring of $N$ cells labelled with a single index $i = 1,2,\ldots, N$ (i.e.\ a line of cells with periodic boundary conditions imposed). As an example, we study the auto-repression Hes/Her model. We assume that each cell signals equally to both its nearest neighbours, yielding the following model equations:
\begin{equation}
   T\ \dot{H}_i = -H_i + h(H_i(t-\tau_h))\ g\left(\frac{H_{i-1}(t-\tau_g) + H_{i+1}(t-\tau_g)}{2}\right),
   \label{eq:in_line_disc}
\end{equation}
where $i$ denotes cell number and the imposition of periodic boundary conditions implies that $H_{-1} = H_N$ and $H_{N+1} = H_1$. Linearising Eq.
(\ref{eq:in_line_disc}) around the uniform steady-state solution
($H_i = H^*$) by expanding $H_i$ as $H_i = H^* + \bar{H}_i
e^{\lambda t}$ yields the following eigenvalue equations:
\begin{equation}
   0 = A \bar{H}_{i-1} + (B - \lambda T) \bar{H}_i + A \bar{H}_{i+1}, \quad i=1,2,\ldots, N,
   \label{eq:eigen_i}
\end{equation}
where
\begin{eqnarray*}
   A(\lambda) & = & -\frac{1}{2} h(H^*) \gamma^g e^{-\lambda \tau_g},\\
   B(\lambda) & = & -1 - g(H^*) \gamma^h e^{-\lambda \tau_h}.
\end{eqnarray*}
Eq.~(\ref{eq:eigen_i}) can be represented in matrix form as:
\begin{widetext}
\begin{equation}
\left( \begin{array}{c}
0 \\
0 \\
0 \\
\\
\vdots \\
\\
0 \end{array} \right) = \left( \begin{array}{ccccccccc}
B - \lambda T & A & 0 & 0 & 0 & \cdots & 0 &  0 & A \\
A & B - \lambda T & A & 0 & 0 & \cdots & 0 & 0 & 0 \\
0 & A & B - \lambda T & A & 0 & \cdots & 0 & 0 & 0 \\
\\
& & & & \cdots & & & & \\
\\
A & 0 & 0 & 0 & 0 & \cdots & 0 & A & B - \lambda T
\end{array} \right)
\left( \begin{array}{c}
\bar{H}_1 \\
\bar{H}_2 \\
\bar{H}_3 \\
\\
\vdots \\
\\
\bar{H}_N \end{array} \right) = \mathbb{M} \left( \begin{array}{c}
\bar{H}_1 \\
\bar{H}_2 \\
\bar{H}_3 \\
\\
\vdots \\
\\
\bar{H}_N \end{array} \right), \label{eq:M}
\end{equation}
\end{widetext}
where it is important to note that $A$ and $B$ are non-polynomial functions of $\lambda$. The eigenvalues $\lambda$ are determined by
$\det \mathbb{M} = 0$.

We first note that the phase relations between adjacent cells in oscillatory solutions can be determined from the form of the eigenvector $(\bar{H}_1, \bar{H}_2, \ldots \bar{H}_N)$ corresponding to each eigenvalue. For any value of $N$, Eq.~(\ref{eq:M}) has an eigenvector $(1, 1, \ldots, 1)$ with an eigenvector determined by the solutions of $T\lambda = B + 2A$. This corresponds to an oscillatory mode where all cells are in-phase. Furthermore, for any even value of $N$, Eq.~(\ref{eq:M}) has an eigenvector $(1, -1, 1, \ldots, 1, -1)$ with an eigenvector determined by the solutions of $T\lambda = B - 2A$. This corresponds to an oscillatory mode where all cells are out-of-phase. These two cases are simple extensions of the dynamics observed in the two cell model.

To illustrate potential extensions to the dynamics observed in the two cell model, we consider the specific case $N = 4$:
\begin{equation}
\mathbb{M} = \left( \begin{array}{ccccccccc}
B - \lambda T & A & 0 & A \\
A & B - \lambda T & A & 0 \\
0 & A & B - \lambda T & A \\
A & 0 & A & B - \lambda T
\end{array} \right),
\label{eq:M_BBBB}
\end{equation}
and
\begin{equation*}
\det \mathbb{M} = (-2 A + T \lambda - B) (2 A + T \lambda - B) (-B + T \lambda)^2.
   \label{eq:eigen_N4}
\end{equation*}
Therefore, the eigenvalues are determined by the solutions of:
\begin{eqnarray}
   T \lambda & = & B \mp 2 A,   \label{eq:eigen_BBBB_BA}\\
   T \lambda & = & B.           \label{eq:eigen_BBBB_B}
\end{eqnarray}
The explicit expression of Eqs. (\ref{eq:eigen_BBBB_BA}) and
(\ref{eq:eigen_BBBB_B}) are found to be the same as the eigenvalue
equations derived for the auto-repression two-Hes/Her model (Eq.
(\ref{eq:eigen_BB})), and for the auto-repression single Hes/Her
model (the minus branch of Eq. (\ref{eq:eigen_AA_no_auto})),
respectively.

\begin{figure}[htbp] 
(a)        \includegraphics[scale=0.4]{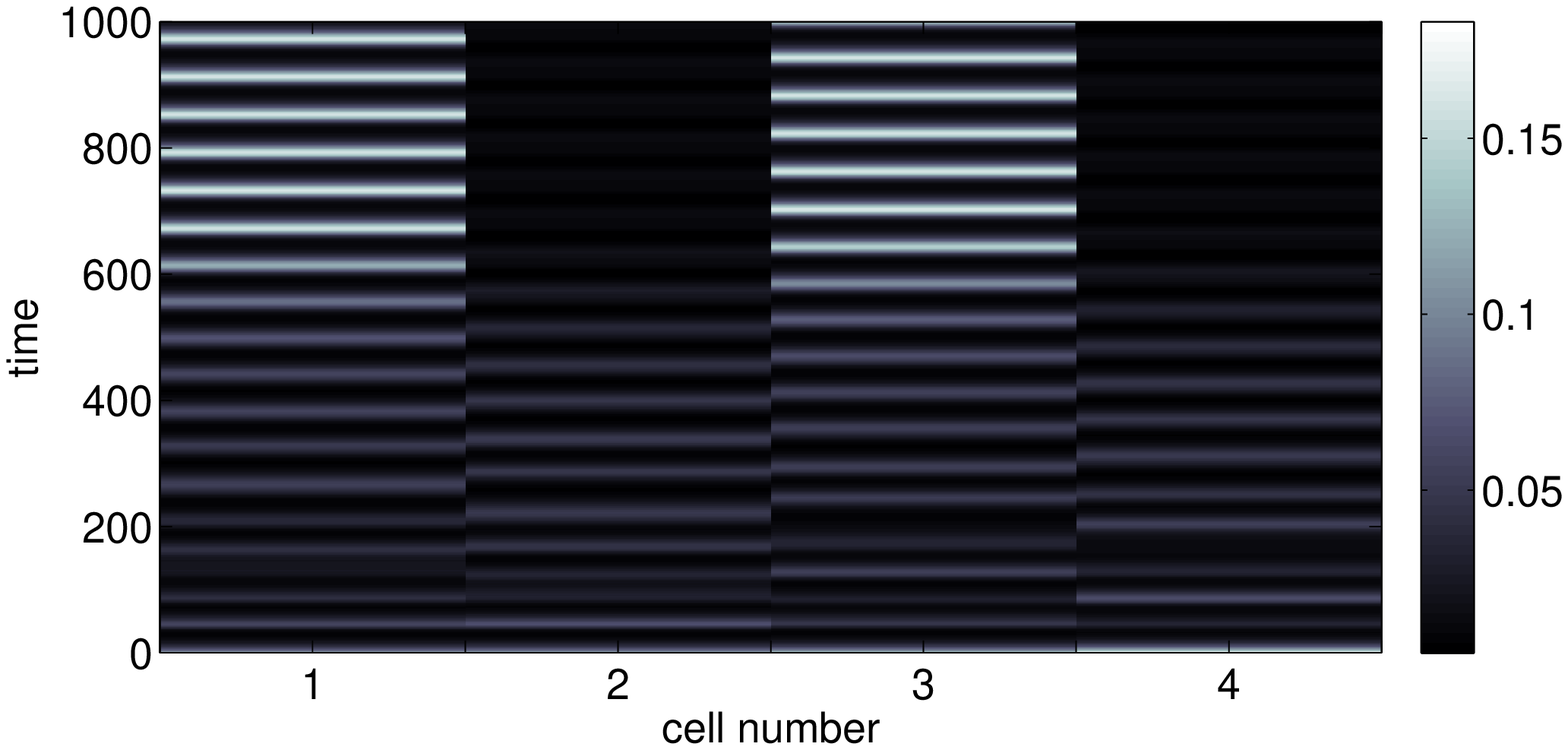}\\ 
(b)        \includegraphics[scale=0.4]{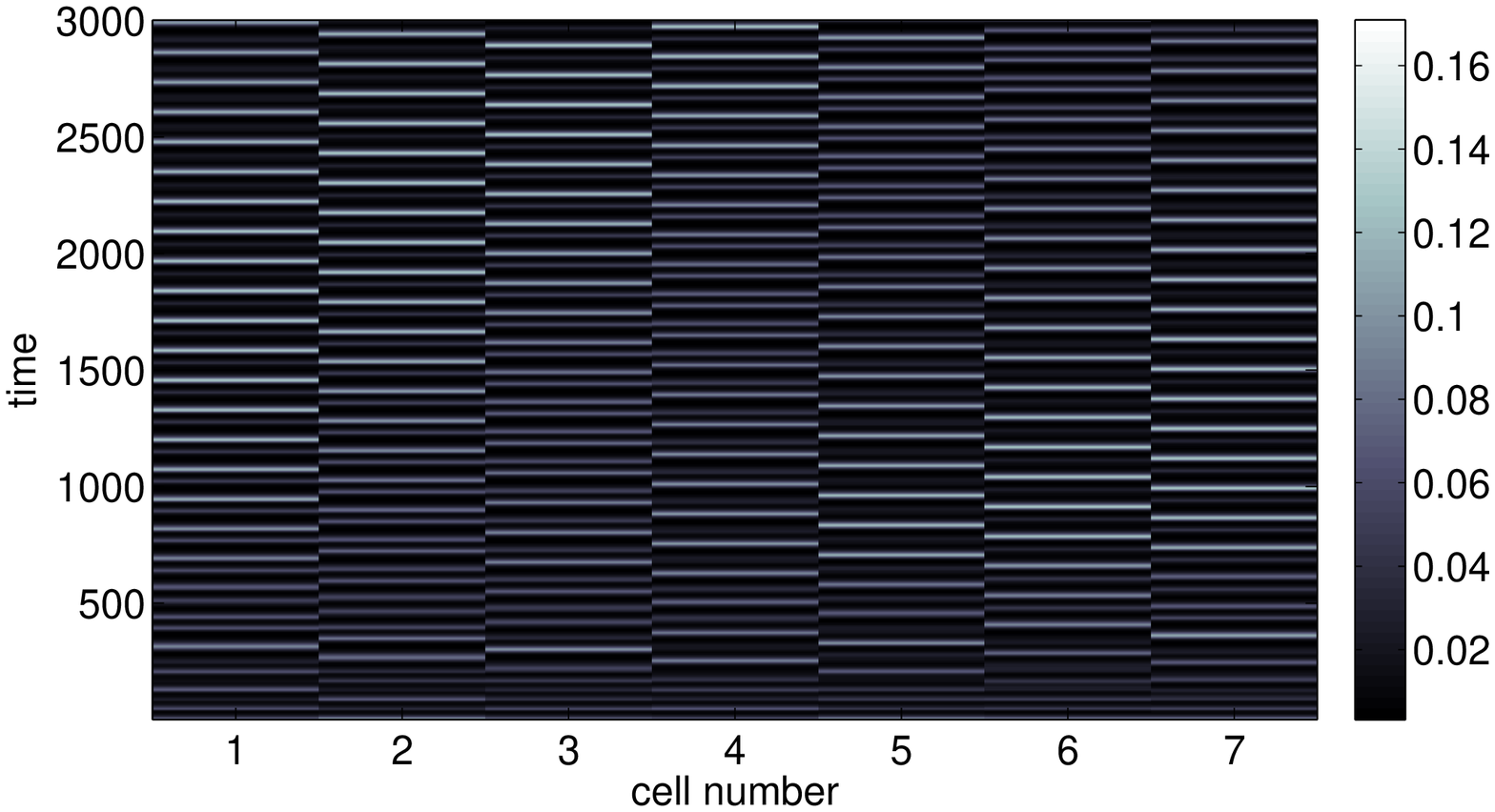}\\ 
       \caption{\label{fig:Ncells}
       Complex oscillatory dynamics in one-dimensional arrays of coupled cells. Simulations of Eqs.~(\ref{eq:in_line_disc}), representing the the auto-repression Hes/Her model in a ring of $N$ cells. Simulations were carried out using the standard parameter values (Table 1) with $\tau_h=20$, $\tau_g=8$ and with initial history $H_i(t) = 0.1 + 0.1\xi_i$ for $-\textrm{max}(\tau_g, \tau_h) \leq t \leq 0$, where the $\xi_i$ are drawn from independent uniform distributions on $[0,1]$ (a) Amplitude death combined with out-of-phase oscillations: $N=4$. (b) Complex dynamics resulting from mode interactions: $N=7$. }
\end{figure}

As above, the phase relations between adjacent cells in oscillatory solutions can be determined from the form of the eigenvector $(\bar{H}_1, \bar{H}_2, \bar{H}_3, \bar{H}_4)$ corresponding to each eigenvalue. From Eq. (\ref{eq:M_BBBB}), the eigenvector on the minus branch of Eq. (\ref{eq:eigen_BBBB_BA}) is found to be $(1, -1, 1, -1)$, representing the out-of-phase oscillatory mode, while the eigenvector on the plus branch is  $(1, 1, 1, 1)$, representing the in-phase mode. In contrast, the eigenvector associated with Eq. (\ref{eq:eigen_BBBB_B}), which is present when $N = 4, 8, 16,\ldots$, is found to be $(1, 0, -1, 0)$, representing amplitude death for every other cell in the array, with the remaining cells being out-of-phase. This is a new dynamical feature that the two-cell model fails to capture. An example of this mode for $N=4$ is shown in Fig.~\ref{fig:Ncells}(a).

For larger values of $N$, interaction between multiple modes can result in more complex oscillatory behaviour, in which waves of phase and amplitude differences can propagate around the ring. Furthermore, the oscillatory profile for each cell is often complex, with multiple peaks per oscillatory period and ``intermittent" oscillations being common. An example for $N=7$ is shown in Fig.~\ref{fig:Ncells}(b). These features are also observed in simulations on regular square and hexagonal two-dimensional arrays of cells (results not shown).

\section{Analysis and reductions of the full Hes/Her--Proneural model} \label{sec:full}
In this section, from the viewpoint of eigenvalue equations, it is
discussed how the full Hes/Her--Proneural model
(Fig.~\ref{fig:motifs}(a)) is related to the auto-activation
Proneural model (Fig.~\ref{fig:motifs}(b)), and to the
auto-repression Hes/Her model (Fig.~\ref{fig:motifs}(c)).

For the uniform steady-state solution ($H_1^*=H_2^*=H^*$,
$A_1^*=A_2^*=A^*$) of the full Hes/Her--Proneural model, the
eigenvalue equation is derived from its model equations
(\ref{eq:4var_1})--(\ref{eq:4var_4}) to be:
\begin{equation}
   (1 + T_H \lambda + a e^{-\lambda \tau_3}) (1 + T_A \lambda - c e^{-\lambda \tau_4})
   = \pm b d e^{-\lambda \tau_{tot}},
   \label{eq:eigen_BABA}
\end{equation}
where $a = f_1(A^*) \gamma^{g_2}$, $b = f_2(A^*) \gamma^{g_1}$, $c =
g_1(H^*) \gamma^{f_2}$, $d = g_2(H^*) \gamma^{f_1}$, and $\tau_{tot}
= \tau_1 + \tau_2$.

First, explicit forms of the neutral angular frequency ($\omega_c$)
and delay ($\tau_{tot}$) are derived for a simple example, where
$T_A = T_H = T$, $\tau_3 = \tau_4 = \tau_a$, and $f_1 = f_2 = g_1 =
g_2$. The last assumption means that $K$ and $\nu$ are respectively
the same all in $f_1$, $f_2$, $g_1$, and $g_2$. Then, from Eqs (9)
and (10), $A^* = H^*$, and consequently $a = c = b = d := \phi$.
Therefore, Eq. (\ref{eq:eigen_BABA}) becomes:
\begin{eqnarray}
   1 + 2 T \lambda + T^2 \lambda^2
       - \phi^2  e^{-2 \lambda \tau_a}
   & = & \pm \phi^2 e^{-\lambda \tau_{tot}}.
   \label{eq:eigen_BABA_symmetric}
\end{eqnarray}
For a neutral solution, Eq.~(\ref{eq:eigen_BABA_symmetric}) becomes:
\begin{equation}
   (1 - T^2 \omega^2 - \phi^2 \cos 2 \omega \tau_a)^2
       + (2 T \omega + \phi^2 \sin 2 \omega \tau_a)^2 = \phi^4,
\end{equation}
\begin{equation}
   \tan \omega \tau_{tot}
   = \mp \frac{2 T \omega + \phi^2 \sin 2 \omega \tau_a}
       {1 - T^2 \omega^2 - \phi^2 \cos 2 \omega \tau_a} := \mp \Delta.
\end{equation}
For the plus branch in Eq.~(\ref{eq:eigen_BABA}):
\begin{equation}
   \tau_{tot}
   = \frac{1}{\omega} \max [\ \tan^{-1} (-\Delta),\ \pi + \tan^{-1} (-\Delta)\ ],
\end{equation}
whereas for the minus branch in Eq.~(\ref{eq:eigen_BABA}):
\begin{equation}
   \tau_{tot}
   = \frac{1}{\omega} \max [\ \tan^{-1} (+\Delta),\ \pi + \tan^{-1} (+\Delta)\ ].
\end{equation}

\begin{figure}[htbp] 
       \includegraphics[scale=0.3]{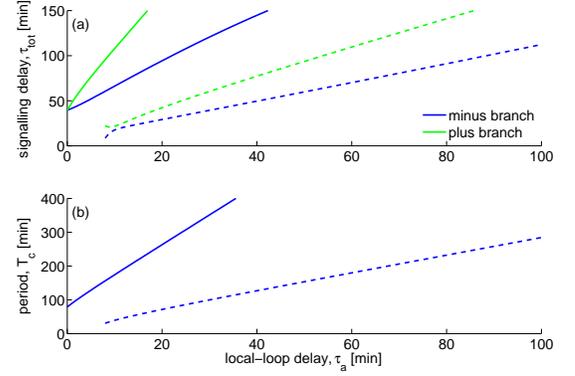} 
       \caption{\label{fig:lnr_BABA} (Color online)
       Neutral (a) intercellular signalling delay and (b) period, associated with the pure imaginary eigenvalues of the full Hes/Her-Proneural model, shown as a function of the local-loop
delay. The eigenvalue equation (Eq.~(\ref{eq:eigen_BABA_symmetric}))
is solved for its minus and plus branches, for the first and the
second largest periods (solid and dashed lines, respectively). In
comparison to Figs.~\ref{fig:lnr_AA} and \ref{fig:lnr_BB}, the
auto-repression two-Hes/Her circuit component is found to be
dominant in this full Hes/Her-Proneural model.}
\end{figure}

Fig.~\ref{fig:lnr_BABA} shows the neutral (a) intercellular
signalling delay and (b) period, as a function of the local-loop
delay ($\tau_a$). The eigenvalue equation
(Eq.~(\ref{eq:eigen_BABA_symmetric})) is solved for its minus and
plus branches, for the first and the second largest periods. In
comparison to Figs.~\ref{fig:lnr_AA} and \ref{fig:lnr_BB}, the
auto-repression two-Hes/Her circuit component is found to be
dominant in this full Hes/Her-Proneural model, although the neutral
values are found to exist for any $\tau_a$, unlike in the
auto-repression two-Hes/Her model, where no pure imaginary solution
can exist for $\tau_f$ less than 5.0483 ($\tau_f$ corresponds to
$\tau_a$ in this section).

In the following, based on the eigenvalue equation of the full
Hes/Her--Proneural model (Eq.~(\ref{eq:eigen_BABA})), we clarify the
conditions under which this full Hes/Her--Proneural model can be
reduced to the auto-activation two-Proneural model
(Fig.~\ref{fig:motifs}(b)), and to the auto-repression two-Hes/Her
model (Fig.~\ref{fig:motifs}(c)).

\subsection{Reduction to the auto-activation two-Proneural model}
The eigenvalue equation of the full Hes/Her--Proneural model
(Eq.~(\ref{eq:eigen_BABA})) is reduced to the eigenvalue equation of
the auto-activation two-Proneural model (Eq.~(\ref{eq:eigen_AA}))
when: [a] $T_H = 0$; [b] $\gamma^{g_2} = 0$; [c] $H^* = A^*$; [d]
$g_2(H^*) \gamma^{f_1} = 1$; and [e] $\tau_1 = 0$. These conditions
must hold irrespective of whether $f_i$ and $g_i$ are Hill functions
or not. If Hill functions $f(x,K,\nu)$ and $g(x,K,\nu)$, where $K$
and $\nu$ are the scaled threshold and the Hill coefficient, are
employed, condition [b] can be met by setting $K$ in $g_2$ to be
$+\infty$. Because this operation makes $g_2(H^*) = 1$, from the
steady-state solution of the full model ($H^* = f_1(A^*)\
g_2(H^*)$), condition [c] is found to require that $f_1$ is not a
Hill function but a linear function: $f_1(x) = x$, whereby
$\gamma^{f_1} = 1$, making all five conditions satisfied. The last
operation ($f_1(x) = x$) means that for a sequence of two functions
$f$ and $g$ to be represented only by $g$, $f$ needs to be $f(x) =
x$. In short, the auto-activation two-Proneural model
(Eqs.~(\ref{eq:model_AA_1}) and (\ref{eq:model_AA_2})) is obtained
by assuming the following conditions on the full Hes/Her--Proneural
model (Eqs.~(\ref{eq:4var_1})--(\ref{eq:4var_4})): the activation
from a proneural protein ($A_{i=1,2}$) to the adjacent Hes/Her
($H_{i=2,1}$) is linear and instantaneous; and Hes/Her is not
associated with an auto-repression loop.

\subsection{Reduction to the auto-repression two-Hes/Her model}
The eigenvalue equation of the full Hes/Her--Proneural model
(Eq.~(\ref{eq:eigen_BABA})) is reduced to the eigenvalue equation of
the auto-repression two-Hes/Her model (Eq. (\ref{eq:eigen_BB}))
when: [a] $T_A = 0$; [b] $\gamma^{f_2} = 0$; [c] $f_1(A^*) =
g_1(H^*)$; [d] $f_2(A^*) \gamma^{f_1}$; [e] $\tau_1 = 0$. These
conditions must hold irrespective of whether $f_i$ and $g_i$ are
Hill functions or not. If Hill functions are employed, condition [b]
can be met by setting $K$ in $f_2$ to be 0. Because this operation
makes $f_2(A^*) = 1$, from the steady-state solution of the full
model ($A^* = g_1(H^*)\ f_2(A^*)$), condition [c] is found to
require that $f_1$ is not a Hill function but a linear function:
$f_1(x) = x$, whereby $\gamma^{f_1} = 1$, making all five
conditions.

\section{General discussion}
The development of multicellular organisms is a process of
sequential and concurrent cell differentiations, the timings of
which must be tightly controlled. Recent experimental studies have
revealed that in the case of vertebrate neural differentiation, cell
differentiation can occur after a transient oscillation in cell
state \cite{ShimojoH08}. Since neural differentiation in vertebrates
occurs over a considerable time interval, these transient
oscillations may play a role in the scheduling of neural
differentiation in relation to other developmental events. The
occurrence of such transient oscillations on the route to neural
differentiation was previously predicted in a simple delay model of
Delta-Notch-medated lateral inhibition \cite{VeflingstadSR05}.

In the present study, we have analysed a more detailed model of the
neural differentiation network. The model has a nested-loop
structure, with intercellular signalling (as mediated by
interactions between DSL ligands and Notch family receptors) coupled
to local cell-autonomous feedback loops. Using a combination of
stability analysis and numerical simulation, we have shown that the
incorporation of local feedback loops has potentially significant
impacts on the dynamics of the signalling network.  In particular,
the time delays in the local feedback loops were found to play a
central role in controlling the behaviour of the whole network:
whether it heads towards differentiation; whether it shows an
oscillation; or whether such an oscillation is sustained or
transient, as well as providing tunability in the amplitude and
period of oscillations.

Many classes of auto-regulatory genes have been identified. Because
these genes are also nodes in larger genetic regulatory networks,
nested feedback loop structures are rather common. The results of
the present study highlight the need for careful examination of the
predictions made by non-delay network models, which include the
\emph{Drosophila} neurogenic model studied by Meir {\em et al.} that
yielded a conclusion that the total system was robust to local
changes to the network circuitry \cite{MeirE02}.

In a more general perspective, a biological system is known to
involve a relatively small number of genes, having numerous features
and functions. Some of the new functions may have been acquired by
the addition of a few local loops to the old conserved ones.
Moreover, because the loops considered in the present study are all
very simple, the outcomes of the present study may have relevance to
non-biological systems as well.

\end{document}